\documentclass[dvips,12pt]{article}
\usepackage{epsfig}%,rotating}
\usepackage{cite}
\def\be{\begin{equation}}
\def\ee{\end{equation}}
\def\bea{\begin{eqnarray}}
\def\eea{\end{eqnarray}}
\def\barr{\begin{array}}
\def\earr{\end{array}}
\def\ra{\rightarrow}

\newcommand{\Z}{{\cal Z}}

\catcode`@=11 % This allows us to modify PLAIN macros.
\def \gsim{\mathrel{\mathpalette\@versim>}}
\def \lsim{\mathrel{\mathpalette\@versim<}}
\def \@versim#1#2{\lower0.4ex\vbox{\baselineskip\z@skip\lineskip\z@skip
     \lineskiplimit\z@\ialign{$\m@th#1\hfil##\hfil$%
     \crcr#2\crcr\sim\crcr}}}
\catcode`@=12 % at signs are no longer letters
\title{
\vspace*{-1.3cm}
\begin{flushright}
\normalsize{
ANL-HEP-PR-02-043\\
EFI-02-93\\
FERMILAB-PUB-02/135-T
  }
\end{flushright}
\vspace{1.5cm}
\Large
\textbf{ Branes and Orbifolds are Opaque
} 
\vspace*{1.0cm}
\author{\large\textbf{Marcela Carena$^a$}, 
\textbf{Tim M.P. Tait$^b$}
and \textbf{C.E.M.~Wagner$^{b,c}$}\\ \\[0.5cm]
$^a$\normalsize\emph{Fermi National Accelerator Laboratory,
P.O. Box 500, Batavia, IL 60510, USA} \\
$^b$\normalsize\emph{HEP Division, Argonne National Laboratory,
9700 Cass Ave.,
Argonne, IL 60439, USA} \\
$^c$\normalsize\emph{Enrico Fermi Institute, Univ. of Chicago, 5640
Ellis Ave., Chicago, IL 60637, USA}}}

\begin{document}
\setcounter{page}{0}
\maketitle
%\date{}
\vspace*{1cm}
\begin{abstract}
We examine localized kinetic terms for
gauge fields which can propagate into compact extra dimensions.  We
find that such terms are generated by radiative corrections in both
theories with matter fields confined to branes and in theories 
imposing orbifold boundary conditions on bulk matter.  In both cases,
the radiative corrections are logarithmically divergent, indicating that from
an effective field theory point of view they cannot be predicted in
terms of other parameters, and should be treated as independent 
leading order parameters of the theory.  Specializing to the five 
dimensional case, we show that these terms may result in gross 
distortions of the Kaluza-Klein gauge field masses,
wave functions, and couplings to brane and bulk matter.
The resulting phenomenological implications 
are discussed.
\end{abstract}

\thispagestyle{empty}
\newpage

%%%%%%%%%%%%%%%%%%%%%%%%%%%%%%%%%%%%%%%%%%%%%%%%%%%%%%%%%%%%%%%
\section{Introduction}
\label{sec:intro}
%%%%%%%%%%%%%%%%%%%%%%%%%%%%%%%%%%%%%%%%%%%%%%%%%%%%%%%%%%%%%%%

Particle physics currently finds itself in the perplexing situation
in which most experimental results conform to the expectations
of the Standard Model (SM), but leave many theoretical
questions unanswered.  For example, the mechanism of electroweak
symmetry breaking is currently unverified, and the mystery as to why the
weak scale is so much smaller than the Planck scale remains unanswered.
The spectrum of quark masses and mixings has been experimentally determined,
and progress is being made on the corresponding quantities for the leptons, 
but no clues as to why the pattern observed shows large hierarchies in
masses and mixings have been established.

Extra compact dimensions allow for novel solutions to these (and other) 
mysteries. By diluting gravity in a (relatively) large volume which gauge 
fields and matter cannot enter, one can lower the fundamental Planck scale
to just above the weak scale, ameliorating the hierarchy problem
\cite{Arkani-Hamed:1998rs}.
There also exist compelling reasons to consider gauge fields which may
propagate into extra dimensions.  Having the gauge fields in the bulk
may allow one to address questions as to why low-scale gravitational effects
do not cause unacceptably fast proton decay \cite{Arkani-Hamed:1999dc},
pursue a geometric origin for the observed spectrum of fermion masses
\cite{Arkani-Hamed:1999dc,Dvali:2000ha,Kaplan:2001ga},
naturally break the electroweak symmetry through strong dynamics
\cite{Cheng:1999bg,Arkani-Hamed:2000hv,He:2001fz},
identify the Higgs as an extra-dimensional component of the gauge field
thus protecting its mass from large corrections
\cite{Hall:2001zb},
achieve gauge coupling unification at high scales
\cite{Dienes:1998vh,Dienes:1998vg}, 
provide a viable dark matter candidate \cite{Servant:2002aq},
and can provide interesting alternatives
to GUT symmetry-breaking and associated problems such as
the Higgs doublet-triplet splitting problem
\cite{Kawamura:2000ev,Hall:2001pg,Hall:2001xr,Hebecker:2001wq,Contino:2001si}.
Provided the compactification scale, related to the size of the extra
dimension by $M_c \equiv 1 / 2 \pi R$, is not much larger than
1 TeV, interesting collider signatures involving production
of Kaluza-Klein (KK) modes of the gauge fields through the scattering
of either brane-localized matter 
\cite{Lykken:1999ms,Dicus:2000hm,McMullen:2001zj}
or bulk matter \cite{Appelquist:2000nn,Rizzo:2001sd,DelAguila:2001pu,Macesanu:2002db,Cheng:2002ab} 
fields may be obtained.

Models with gauge fields in more than four dimensions are not 
renormalizable in the classic sense, and must be regarded as effective
theories which break down at some scale $\Lambda$.  In fact, because of the
rapid classical
evolution of the coupling constant in more than four dimensions, the
gauge coupling becomes strong at energy scales on the order of ten times
the compactification scale, and thus the scale $\Lambda$ is expected to
be relatively close to $M_c$.  Since the nature of the UV completion
is unknown, these theories should be understood as an expansion in
the energy of the process at hand, with effects of the unknown physics
beyond $\Lambda$ reflected in the (infinite number of) undetermined 
coefficients which must be treated as theoretical inputs.  The theory
can be predictive at energies much below $\Lambda$, when effects of
order $E^n / \Lambda^n$ may be neglected, and only a finite number
of terms in the effective Lagrangian contribute to measurable
quantities.

In this article we attempt to rigorously treat gauge fields which
can propagate in compact extra dimensions from an effective field theory
point of view.  We find that in addition to the bulk kinetic terms for
the gauge field generally considered in the literature, there is 
also a kinetic term for the gauge field localized on branes or
at the boundaries of an orbifolded compact space.  
Such a term was recently considered by Dvali, Gabadadze, and Shifman
\cite{Dvali:2000rx}.
The original motivation was to have the fifth dimension {\em infinite}
in size, with the brane term allowing one to recover four dimensional
behavior at short distances, but in this article we will show that,
as happens in the analogous case of gravity \cite{Carena:2001xy},
it has interesting implications for compact spaces as well.
The brane kinetic term is not suppressed by any power of 
$\Lambda$ compared to the 5d couplings which result in the apparently
renormalizable 4d interactions at low energies, 
and is consistent with all symmetries of the theory.  
Thus, from general renormalizability arguments, one expects that the term 
{\em must} be included in any consistent description of the 
theory\footnote{For an example of a string theory model in which gauge kinetic
terms on boundaries occur at tree level with calculable magnitude,
see \cite{Antoniadis:1999ge,Ibanez:1998qp}.}.
In fact, one can show from explicit computation
that such a term is required to cancel divergences in the 
five dimensional (5d) theory.
Thus, its magnitude should be treated as an input to the theory, and
one might expect it to be sizable.  

In this article, while we have chosen to illustrate the physics with the 
specific example of bulk gauge fields, we recognize that by no means is
this the only possibility.  {\em Any} bulk field will experience 
renormalizations on branes or boundaries of the type we are describing.
Previous work 
\cite{Carena:2001xy,Csaki:2000fc,Csaki:2000pp,Dvali:2000hr,Dvali:2000xg}
has focused on the case of gravity in various background geometries and 
numbers of dimensions, as is motivated by solutions of the hierarchy
problem.  We choose to work with gauge theories in five dimensions because,
aside from being well-motivated for the reasons outlined above, they
are under better theoretical control than theories with quantum gravity
or larger numbers of extra dimensions.  We find that some of the qualitative
results seen in the gravitational case, such as the appearance of
``collective'' Kaluza-Klein modes with small masses and strong couplings,
may also be explored in our framework.

The article is organized as follows.  In Section~\ref{sec:brane}
we briefly discuss the existence of such a brane kinetic term, 
and argue that from an effective theory point of view it should be included.
In Section~\ref{sec:kkmodes}, we compute the resulting spectrum of
KK modes of the gauge fields and examine the masses and couplings
to brane and bulk fields.  In Section~\ref{sec:pheno} we examine some
of the phenomenological implications of the modifications
to masses and couplings.  
We reserve Section~\ref{sec:conclusion} for our conclusions.

%%%%%%%%%%%%%%%%%%%%%%%%%%%%%%%%%%%%%%%%%%%%%%%%%%%%%%%%%%%%%%%
\section{Framework and Brane Kinetic Terms}
\label{sec:brane}
%%%%%%%%%%%%%%%%%%%%%%%%%%%%%%%%%%%%%%%%%%%%%%%%%%%%%%%%%%%%%%%

We now discuss the existence of local gauge kinetic terms from the
point of view of effective field theory.  To illustrate our discussion,
we consider a five dimensional (5d) theory of gauge fields
${\cal A}^M$,
\bea
\label{eq:s5d}
S &=& \int d^5 x \left\{
-\frac{1}{4 g_5^2} {\cal F}^{M N} {\cal F}_{M N} 
- \delta (x_5) \frac{1}{4 g_a^2} {\cal F}^{\mu \nu} {\cal F}_{\mu \nu} 
\right\} ,
\eea
where capital latin letters refer to the full 5d coordinates,
$M = 0, 1,2 ,3, 5$, and lower case greek letters refer only to
the four uncompactified dimensions, $\mu = 0, 1, 2, 3$.  Note that
written this way, the bulk gauge coupling $g_5$ has mass dimension
$-1/2$ and the gauge field ${\cal A}^M(x^\mu, x_5)$ has dimension $1$.
The brane coupling $g_a$ is dimensionless and characterizes the ``opacity''
of the brane.  Analyses which neglect the brane term
($1/ g^2_a \ra 0$) can be understood as the ``transparent
brane'' approximation.  One may rescale the bulk
term to its canonical normalization by absorbing $1/g_5$ into
${\cal A}_M$, in which case the brane term has a coefficient with
dimensions of length, $r_c = g_5^2 / g_a^2$ and ${\cal A}_M$
has dimension $3/2$, as usual for a boson in five dimensions.
${\cal F}$ is the usual 
field-strength functional of the gauge fields,
\bea
{\cal F}^a_{M N} &=& \partial_M {\cal A}^a_N - \partial_N {\cal A}^a_M 
+ f^{abc} {\cal A}^b_M {\cal A}^c_N ,
\eea
for a non-Abelian Yang-Mills theory, with the final term omitted
in the Abelian case.  We generally omit the group index on the gauge
fields wherever we may do so without confusion.

The 5th coordinate $x_5$ corresponds to a compactified dimension
$S^1 / \Z_2$, with $-\pi R \leq x_5 \leq \pi R$.  Under the orbifold
$\Z_2$, the points $-x_5$ and $x_5$ are identified, and
the fields transform as,
\bea
{\cal A}^\mu (x^\mu, -x_5) & = & \:\:\: {\cal A}^\mu (x^\mu, x_5) \\
{\cal A}^5   (x^\mu, -x_5) & = & -  {\cal A}^5 (x^\mu, x_5) \nonumber .
\eea
The action and orbifold are compatible with the 5d subset of gauge 
transformations (${\cal A}^M \ra {\cal A}^M - \partial^M \lambda(x^M)$
for a U(1) theory) with transformation function $\lambda(x^M)$ 
chosen to be an even function of $x_5$.  

It is important to note that we have added only the four-dimensional
part of the gauge field kinetic term on the brane.  If our theory was invariant
under the full 5d set of gauge and Lorentz transformations, these symmetries
would have forced us to include the full 5d gauge kinetic term.  However,
the 5d Lorentz invariance is broken firstly by the fact that one of the 
dimensions is compact, secondly by the orbifold boundary conditions, and
finally by choosing $x_5=0$ as a special point with different physics from
the rest of the extra dimension.  As discussed above, five dimensional gauge
invariance is similarly present only in a restricted sense.
Thus, one could also consider including
the remaining terms on the brane with a different coefficient,
\bea
\int dx_5 \: \delta (x_5)  \left\{  \frac{1}{2 \tilde{g}_a^2} \left[
\partial_\mu {\cal A}_5  \; \partial^\mu {\cal A}_5
- 2 \: \partial_\mu {\cal A}_5 \; \partial_5 {\cal A}_\mu
+ \partial_5 {\cal A}_\mu  \; \partial_5 {\cal A}_\mu \right]
+ {\rm Interactions} \right\}.
\label{eq:otherterms}
\eea
In the thin brane approximation under which we work, all of these terms
may be neglected.  The first two terms vanish because the orbifold
boundary conditions require ${\cal A}_5$ to vanish at the orbifold fixed points
(and in fact we will impose ${\cal A}_5=0$ everywhere as a convenient
gauge choice).  The last term is somewhat more subtle.  Naively
the orbifold conditions seem to require $\partial_5 {\cal A}_\mu$,
as an odd function of $x_5$, to vanish at the fixed points. However
as we will see below the effect of the brane term is to force the slope
of the KK wave functions to be discontinuous at $x_5=0$, 
implying the derivative
is not well defined in the thin brane approximation.  However, it is
clear that when the derivative is understood in terms of
the difference between the wave function of ${\cal A}_\mu$ around
$x_5=0$, the term vanishes.

One can attempt to consider this problem more carefully by introducing
a finite brane thickness.  For example, one can replace the 
$\delta$-function with any smooth function sharply peaked about the 
orbifold fixed point.  In a ``fat brane'' model, which represents the 
brane and its attendant localized fermions as a scalar field whose VEV has a 
domain wall profile along the extra dimension, this function is related
to the scalar potential which generates the domain wall, and the brane
width and shape can be adjusted.  
This finite thickness for the brane will smooth
our KK wave function solutions such that their derivatives will become
well-defined, and in fact the terms in Equation~(\ref{eq:otherterms}) 
will vanish at $x_5=0$.  However, the terms still have some non-zero 
effect in the region away from $x_5=0$, but still inside the brane.  Thus,
their effect must be proportional to the thickness of the brane, and
we are justified in dropping the terms of Equation~(\ref{eq:otherterms})
in the limit in which we treat the brane as infinitely thin.

\subsection{Branes are Opaque}

Many theories in which gauge fields exist in extra compact dimensions
introduce submanifolds, or branes on which fields may be confined.  A simple
application is to have chiral fermions living on a 3-brane.  This allows one
to have an effective 4d theory which is chiral, despite the fact that five
(or more) dimensional theories generally produce mirror fermions after
compactification to 4d, and thus are vector-like.

The existence of a brane violates the 5d Poincare invariance, and thus
one would generically expect terms living on the brane would be invariant
only under the 4d Poincare invariance of the brane itself.  Thus, it would
be quite plausible to consider a separate gauge kinetic term on the brane
at tree level.  In fact,
the existence of charged matter on the brane {\em demands} such a term.
Loops of the brane matter fields result in $\log$ divergent
contributions to the gauge field 2-point function, localized on the brane
itself \cite{Dvali:2000rx}.
In this case, if the brane is approximated as infinitely thin
the computation becomes effectively four-dimensional because the fields
running around in the loops are four dimensional.  In fact, the
log divergence is nothing more than the familiar renormalization of the
gauge coupling by the brane fields.

The cancellation of the divergence invokes a local term in $x_5$
of the form of the gauge field kinetic term, and indicates that the bare
theory without such a term is inconsistent.  After canceling the divergence,
what is left behind is a term whose coefficient cannot be computed in
terms of other quantities of the theory, but must instead be determined
by experiment.  
As usual, a $\log$ term appears in conjunction with the divergence,
and its resummation dictates that even if one were to imagine a UV
completion which resulted in the local term being zero at some energy
scale, a non-zero term will evolve at other energy scales through 
renormalization group evolution.
One can, of course, choose the coefficient $1/g_a^2$ to be very small, 
but unless one can derive the small value of $1/g_a^2$ within the
framework of a more fundamental theory, this choice can be regarded as
a fine-tuning.

One may also invoke the fat brane picture, allowing the brane to 
have a non-zero thickness corresponding to the width
of the transition region between the two limiting values of the VEV.
In this case, fermions can be localized with wave functions whose
widths are related to the thickness of the fat brane.  Loops of such
fermions will also lead to localized renormalization of the gauge
fields, but now with a profile proportional to the fermion wave
functions, and not to the $\delta$-function one obtains
in the thin brane case.  The detailed shape of the local term is thus 
model-dependent in general fat brane cases.

\subsection{Orbifolds are Opaque}

It is somewhat counter-intuitive that local terms exist for orbifold
theories even in the absence of localized fields.  However, in theories
with an orbifold, the identification of $x_5$ with $-x_5$ indicates that
the sign of momentum along the fifth dimension is not meaningful, and
singles out the orbifold fixed points as special points where translation
invariance is violated.  These effects may be cast into a particularly
convenient form by using technology developed in \cite{Georgi:2000ks},
in which one writes the bulk fields obeying orbifold boundary conditions
as a combination of fields which are unconstrained.  So, for example, a
5d bulk scalar $\Phi$ which is odd under the orbifold $\Z_2$ is written,
\bea
\Phi(x^\mu, x_5) & = & \frac{1}{2} 
\left[ 
\phi(x^\mu, x_5) - \phi(x^\mu, -x_5) \right]
\eea
where $\phi(x^\mu, x_5)$ is a 5d scalar field without orbifold 
boundary conditions, and thus may be treated conventionally.  $\Phi$,
by construction, obeys the orbifold boundary conditions.  The
$\Phi$ propagator in momentum space now contains terms which flip the
sign of the momentum in the extra dimension,
\bea
\langle \Phi \Phi^* \rangle ( q, q^\prime ) & = &
\frac{i}{2} \left\{
  \frac{\delta_{q_5, q^\prime_5} - \delta_{-q_5, q^\prime_5}}{q^2 - q_5^2}
\right\} \delta^4 (q_\mu - q^\prime_\mu) .
\eea
The first term conserves $q_5$ whereas the second induces the sign flip.  If
this scalar is now coupled to a bulk gauge field
${\cal A}^M$, its loop contributions to the
gauge field two-point function will also contain a term which conserves
the gauge field momentum, and a term which conserves its magnitude
but flips its sign \cite{Georgi:2000ks}.  Transforming back to position
space, one has the operator,
\bea
-\frac{r_c}{4} {\cal F}^{\mu \nu} {\cal F}_{\mu \nu} 
\left[ \delta(x_5) + \delta(x_5 - \pi R) \right]
\eea
where $r_c$ contains the loop integrals, and is log divergent.  Again,
this signals that the term is in actuality tree-level, and all we
have divined is the running from the cut-off scale to the energy
scales of interest to us.  Therefore,
even universal extra dimensions, with no fields living on the boundaries,
will generally have kinetic terms which do live on the boundaries.  Note that
the same term is induced on both boundaries, which is important if a
KK parity is to be a self-consistent symmetry of the low energy dynamics.

\subsection{Naive Dimensional Analysis}

While the effective field theory perspective strictly demands the
coefficient of the brane kinetic term to be a free parameter of the
theory, it is interesting to see how large one might expect this term
to be if one makes further assumptions.  In particular, naive
dimensional analysis (NDA) determines the size of various couplings under
the assumption that all couplings are strong at the scale $\Lambda$
\cite{Manohar:1983md}.  While NDA estimates are interesting (and sometimes
useful in order to judge the applicability of perturbation theory), we
do not wish to consider them as predictions -- we take the more practical
view that the brane kinetic terms are remnants of the unknown physics
beyond the cut-off, and must be included irrespective of their size
in any valid effective field theory description.

The techniques of \cite{Chacko:1999hg} allow us to simply determine
the values of the couplings $g_5$ and $r_c$ at $\Lambda$, and we may use
the renormalization group to examine their magnitudes at other energy
scales of interest. The NDA estimate for $r_c$ at $\Lambda$ is given by,
\bea
r_c & \sim & \frac{6 \pi}{\Lambda} ,
\eea
and thus $r_c / R \sim 6 \pi M_c / \Lambda$.  If $\Lambda$ is 
as low as roughly 20 times $M_c$, we have $r_c / R \sim 1$.
At lower energy scales $r_c$ will receive additional logarithmic
corrections under the renormalization group.  Since these corrections
are suppressed by loop factors, they can be considered subdominant 
corrections to the NDA estimates.  Finally, we note that in 
Ref.~\cite{Ponton:2001hq}, theories with large brane kinetic terms (relative 
to their natural scale, $\Lambda$) were found to be perturbatively consistent.
Thus, there is nothing problematic with considering them to be large.

%%%%%%%%%%%%%%%%%%%%%%%%%%%%%%%%%%%%%%%%%%%%%%%%%%%%%%%%%%%%%%%
\section{Kaluza-Klein Decomposition}
\label{sec:kkmodes}
%%%%%%%%%%%%%%%%%%%%%%%%%%%%%%%%%%%%%%%%%%%%%%%%%%%%%%%%%%%%%%%

We now derive the KK decomposition for the gauge fields in the presence
of the brane kinetic term.  Before starting, it is worthwhile to recall
the results for transparent branes ($r_c \ra 0$).  In the
transparent brane case, the
action Equation~(\ref{eq:s5d}) can be decomposed into,
\bea
\frac{1}{g_5^2} \int d x_5 \left\{
-\frac{1}{4} {\cal F}^{\mu \nu} {\cal F}_{\mu \nu} 
+ \frac{1}{2} \partial_5 {\cal A}_\mu  \; \partial_5 {\cal A}^\mu  \right\} ,
\eea
where we have chosen a gauge in which ${\cal A}_5=0$, corresponding
to a unitary gauge in which the fifth components of the gauge field are 
eaten by the 4d components to provide longitudinal degrees of freedom to the
massive modes \cite{SekharChivukula:2001hz}.
We expand the gauge field in a KK tower,
\bea
{\cal A}^\mu (x^\mu, x_5) = \sum_n \; f_n (x_5) \; A^\mu_n (x^\mu)
\label{eq:kkdecomp}
\eea
where $f_n(x_5)$ is a set of complete functions which we choose by
requiring the KK masses to be diagonal,
\bea
\partial_5^2 f_n(x_5) = -m_n^2 \; f_n(x_5) .
\eea
The solution to this equation consistent with the orbifold boundary
conditions are cosines, with frequencies (masses) $m_n = n/R$,
$n=0,1,2,3...$.  There is a zero mode, whose wave function is a constant
in $x_5$, and thus properly normalized is $1/\sqrt{2 \pi R}$.  
The normalization for the cosine functions is given by $1/\sqrt{\pi R}$,
and this difference in normalization results in a $\sqrt{2}$
enhancement of the KK gauge boson coupling to brane matter compared to the
coupling of the gauge boson zero mode ({\em i.e.,} see 
ref.~\cite{Rizzo:1999br}).

\subsection{One Opaque Brane}

In this section we derive the KK decomposition for a five dimensional 
theory in the presence of a single non-zero localized gauge kinetic term.
For small choices of $g_5^2 / g_a^2$, the brane term is a perturbation on
the KK spectrum, introducing a small amount of mixing between the 
KK modes of various levels.  For 
$g_5^2 / g_a^2 \sim R$, these mixing effects drastically affect the KK 
decomposition, and one is no longer justified in treating the brane term
as a perturbation, but should instead include its effect on the KK spectrum
{\em ab initio}.  

In the presence of the brane term, the KK decomposition will be diagonal
if the wave functions $f_n(x_5)$ satisfy,
\bea
\frac{1}{g_5^2} 
\int dx_5 \: \left[ 1 + r_c \, \delta (x_5) \right] \; f_n (x_5) \; f_m (x_5) 
& = & Z_n \, \delta_{n m} , \nonumber \\
\frac{1}{g_5^2} 
\int dx_5 \:  f^\prime_n (x_5) \; f^\prime_m (x_5) & = & 
Z_n \, m^2_n \, \delta_{n m} ,
\label{eq:diagonal}
\eea
where the prime represents partial differentiation with respect to $x_5$.
In order to solve these equations
simultaneously, we follow a variant of the procedure 
used in Ref.~\cite{Dvali:2001gm} to handle scalar fields.  
We begin with the relevant 5d linearized equation 
of motion for the gauge field,
\bea
  \partial_M \partial^M {\cal A}_\mu
- \partial_\mu \left( \partial_M {\cal A}^M \right)
+ r_c \; \delta (x_5) \left\{
  \partial_\nu \partial^\nu {\cal A}_\mu
- \partial_\mu \left( \partial_\nu {\cal A}^\nu \right) \right\}
= 0
\eea
where we have dropped the group index on ${\cal A}$ for convenience.
We now expand the gauge field in a KK tower as in 
Equation~(\ref{eq:kkdecomp}), and determine the $f_n(x_5)$ 
by requiring the $A^\mu_n$ to satisfy the linearized equation of motion of
a 4d massive gauge field,
\bea
  \partial_\nu \partial^\nu A_n^\mu
- \partial^\mu \left( \partial_\nu A_n^\nu \right)
+ m_n^2 A_n^\mu = 0 .
\eea
This procedure becomes particularly simple if we make the 5d gauge
choice ${\cal A}^5 = 0$.
In that case one obtains the equation for the $f_n$:
\bea
\left[ \partial_5^2 + m_n^2 + r_c \; m_n^2 \; \delta (x_5) \right] f_n = 0.
\eea
This equation, which embodies the diagonalization conditions in
Equation~(\ref{eq:diagonal}),
is the same as the equation found in \cite{Dvali:2001gm} for a 
scalar field.  It bears a strong resemblance to the 
nonrelativistic Schr\"odinger equation for a $\delta$-function 
potential whose strength is energy-dependent, and its spectrum
is thus guaranteed to have real eigenvalues.
Away from $x_5 = 0$, the solutions are sums of
sine and cosine functions.  We thus write solutions piece-wise
in the regions $x_5 < 0$ and $x_5 > 0$ and 
impose periodicity and continuity at $x_5 = 0$,
\bea
f_n (x_5 - 2 \pi R) &=& f_n(x_5) \\
f_n ( 0^+ ) &=& f_n ( 0^- ) \nonumber \\
f^\prime_n ( 0^+ ) - f^\prime_n ( 0^- ) &=& - r_c m_n^2 f_n( 0 ),
\nonumber
\eea
where $0^+$ and $0^-$ denote the limit as $x_5$ approaches zero from 
above or below.

\begin{figure}[t]
\vspace*{-1.cm}
\centerline{
\epsfxsize=12.0cm\epsfysize=12.0cm
                     \epsfbox{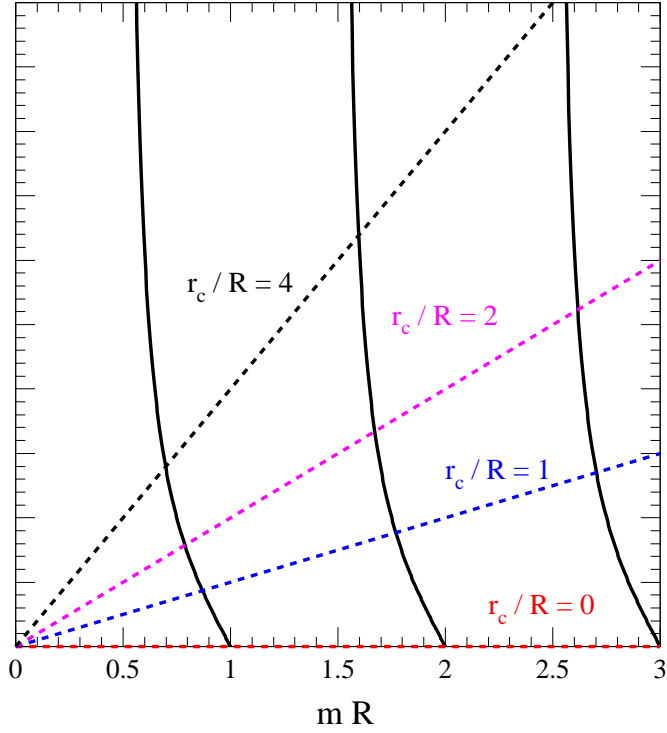}
}
\caption{Graphical solution of the eigenmass equation, 
$-\tan[\pi m R] = (r_c / 2 R) \times m R$ for several values of $r_c / R$.}
\label{fig:solution}
\end{figure}

The resulting solutions have quantized masses which are solutions
of the transcendental equation,
\bea
\label{eq:eigenmasses} 
\frac{r_c \; m_n}{2} & = & -\tan [ \pi \; R \; m_n ]
\hspace*{2cm} \left( m_n \geq 0 \right) \: , 
\eea
which may be solved graphically as in Figure~\ref{fig:solution}.
The corresponding wave functions are,
\bea
\label{eq:eigenmodes}
f_n (x_5) = {\cal N}_n \left\{ 
\begin{array}{lcr}
\cos [m_n x_5 ] + (m_n r_c / 2) \sin [m_n x_5] & & x_5 < 0 \\[0.2cm]
\cos [m_n x_5 ] - (m_n r_c / 2) \sin [m_n x_5] & & x_5 \geq 0
\end{array}
\right. .
\eea
We define the constant ${\cal N}_n$ by normalizing $f_n$ such that,
\bea
\int^{+\pi R}_{-\pi R} \, dx_5 \: f^2_n(x_5) & = & 1 ,
\eea
which results in,
\bea
\frac{1}{{\cal N}_n^2} & = & \pi R \left( 1 + \frac{m_n^2 \, r_c^2}{4}
- \frac{r_c}{2 \pi R} \right) ,
\eea
for $n \geq 1$ and ${\cal N}_0 = 1 / \sqrt{2 \pi R}$.
Inserting this KK decomposition into our original 5d action, 
Equation~(\ref{eq:s5d}), and performing the integration over $x_5$
results in kinetic terms for the gauge fields which are diagonal,
\bea
{\cal L}^0_4 & = &
\sum_n \left\{ -\frac{1}{4} \, Z_n \, 
\left( \partial_\mu A^n_\nu - \partial_\nu A^n_\mu \right)
\left( \partial^\mu A_n^\nu - \partial^\nu A_n^\mu \right)
+ Z_n \; \frac{m_n^2}{2} A_n^\mu A^n_\mu
\right\}
\eea
where $Z_n$ is a normalization factor with dimensions of mass.
This equation is consistent with Equation~(\ref{eq:diagonal}),
indicating that we have successfully diagonalized the KK decomposition.

\begin{figure}[t]
\vspace*{-1.cm}
\centerline{ \hspace*{1cm}
\epsfxsize=10.0cm\epsfysize=10.0cm
                     \epsfbox{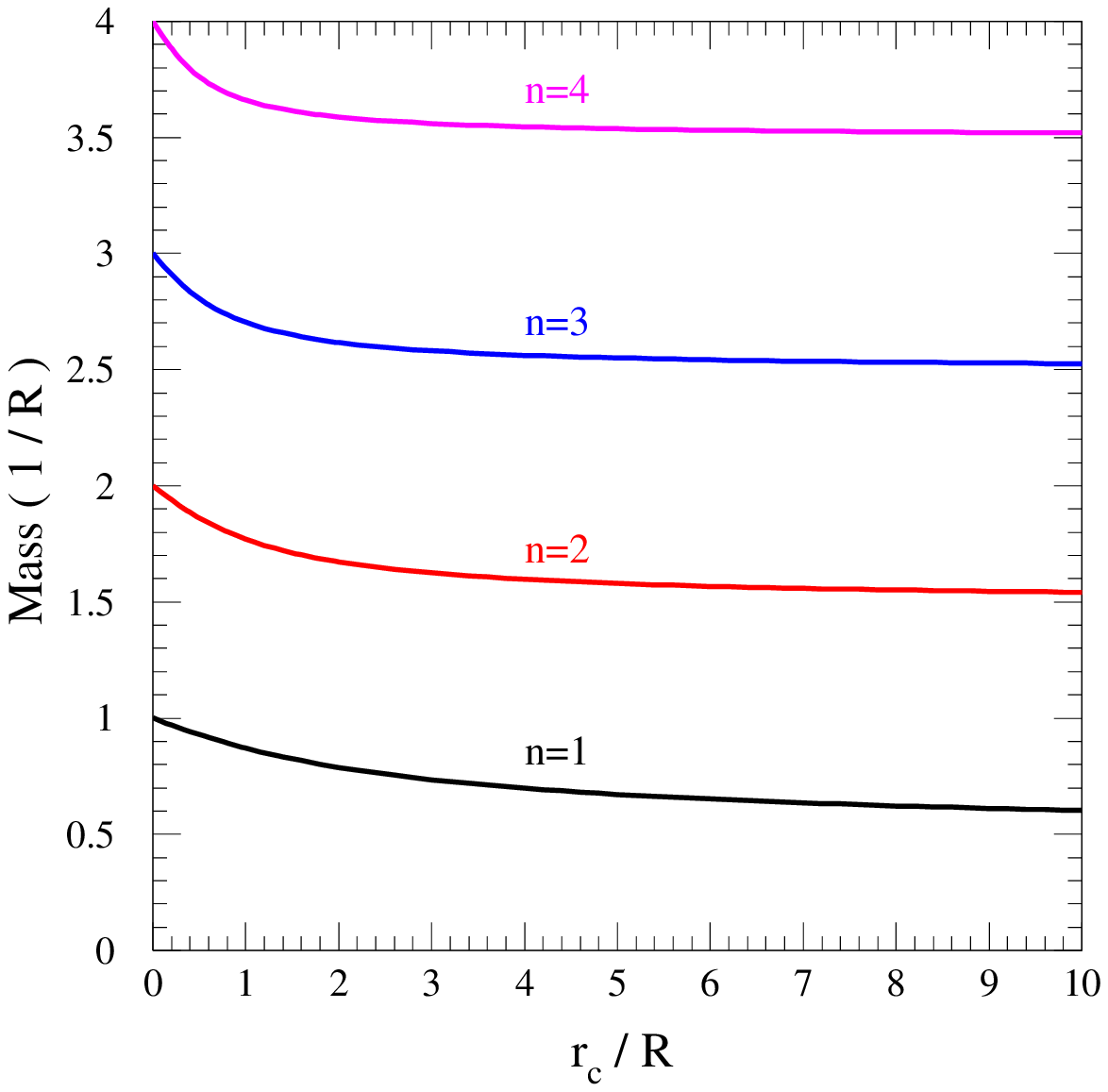} \hspace*{-1.5cm}
\epsfxsize=10.0cm\epsfysize=10.0cm
                     \epsfbox{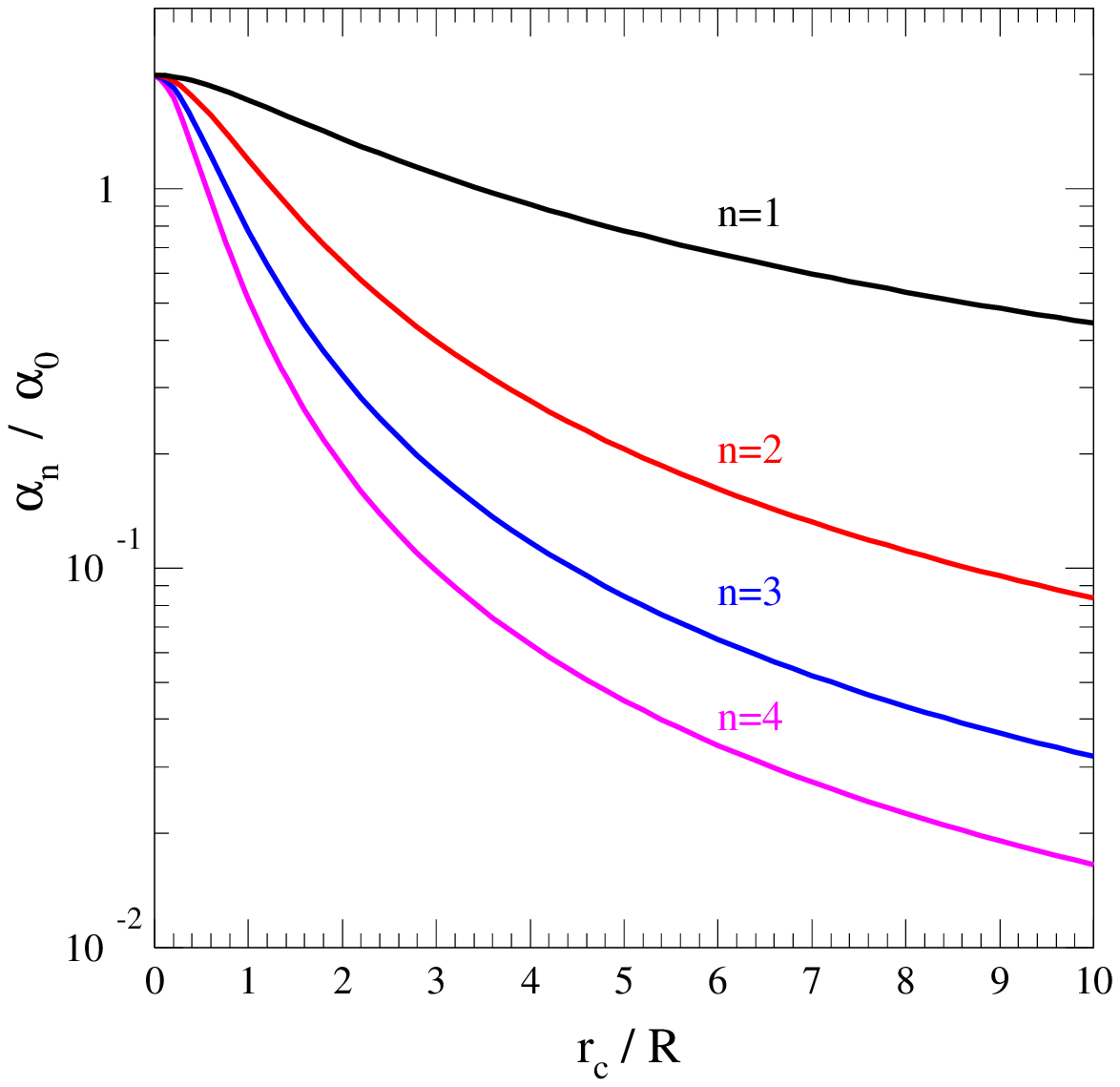}
}
\caption{The $n=1,2,3,4$ (bottom to top) KK mode masses in units of $1/R$
and couplings relative to the zero mode coupling 
for the one-brane case, as a function of $r_c / R$.}
\label{fig:masses}
\end{figure}

Note that Eq.~(\ref{eq:eigenmasses}) always has a solution for $m_n =0$,
and that the corresponding $f_0(x_5)$ is always a constant.  Thus, there
is always a zero mode gauge field whose profile does not depend on the 
extra dimension.
In the limit $r_c \ra 0$, in which the brane kinetic
term is negligible, we reproduce the standard KK spectrum with masses
$n / R$.  In Figure~\ref{fig:masses}, we present the
masses of the first four KK modes as a function of $r_c/R$.
Clearly, for $r_c / R \sim 1$, the spectrum shows some 
distortion in the spacing between the lowest modes.  
For any $r_c/R$, the higher modes asymptote 
to equal spacing of $1/R$ as expected, though the spectrum still shows
an over-all shift dependent on $r_c / R$.  For $r_c \gg R$, the
masses asymptotically approach $n / 2 R$.

It is also instructive to examine the couplings of the KK tower to
various types of fields, either confined to branes or living in the bulk.
Some representative interaction terms in the 5d theory are,
\bea
{\cal L} & = & \int dx_5 \left\{
\delta (x_5 - x_\psi) \; 
\left[ \overline{\psi} {\cal A}_\mu \gamma^\mu \psi \right]
+ \left( \frac{1}{g_5^2} + \frac{\delta(x_5)}{g_a^2} \right)
\left[ 2 (\partial_\mu {\cal A}^a_\nu - \partial_\nu {\cal A}^a_\mu )
f^{abc} {\cal A}_b^\mu {\cal A}_c^\nu  \right]
\right. \nonumber \\
& & \hspace*{2 cm} \left. + 
\left( \frac{1}{g_5^2} + \frac{\delta(x_5)}{g_a^2} \right)
\left[ f^{abc} f^{ade} {\cal A}_b^\mu {\cal A}_c^\nu 
{\cal A}^d_\mu {\cal A}^e_\nu \right]
\right\} .
\eea
The first term represents coupling to a fermion on a brane at $x_\psi$
(for a bulk fermion mode with wave function $f_{\psi}(x_5)$ one
replaces $\delta (x_5 - x_\psi) \ra |f_{\psi}(x_5)|^2$) and
the later two terms are the interactions among the bulk gauge fields
for a non-Abelian theory.  In order to derive the effective interactions
between various KK modes, one inserts the KK decomposition into this equation,
and then rescales each $A^n$ by $Z_n^{-1/2}$ in order to canonically normalize
its kinetic terms.
Given our convention to normalize the $f_n(x_5)$, the
$n$-mode gauge field has $Z_n$,
\bea
Z_n & = & \left( \frac{1}{g_5^2} + \frac{f^2_n(0)}{g_a^2} \right) ,
\eea
where $f_n(0) = {\cal N}_n$ is the wave function of the $n$th mode evaluated at
the origin.

\begin{figure}[t]
\vspace*{-1.cm}
\centerline{ \hspace*{1cm}
\epsfxsize=10.0cm\epsfysize=10.0cm
                     \epsfbox{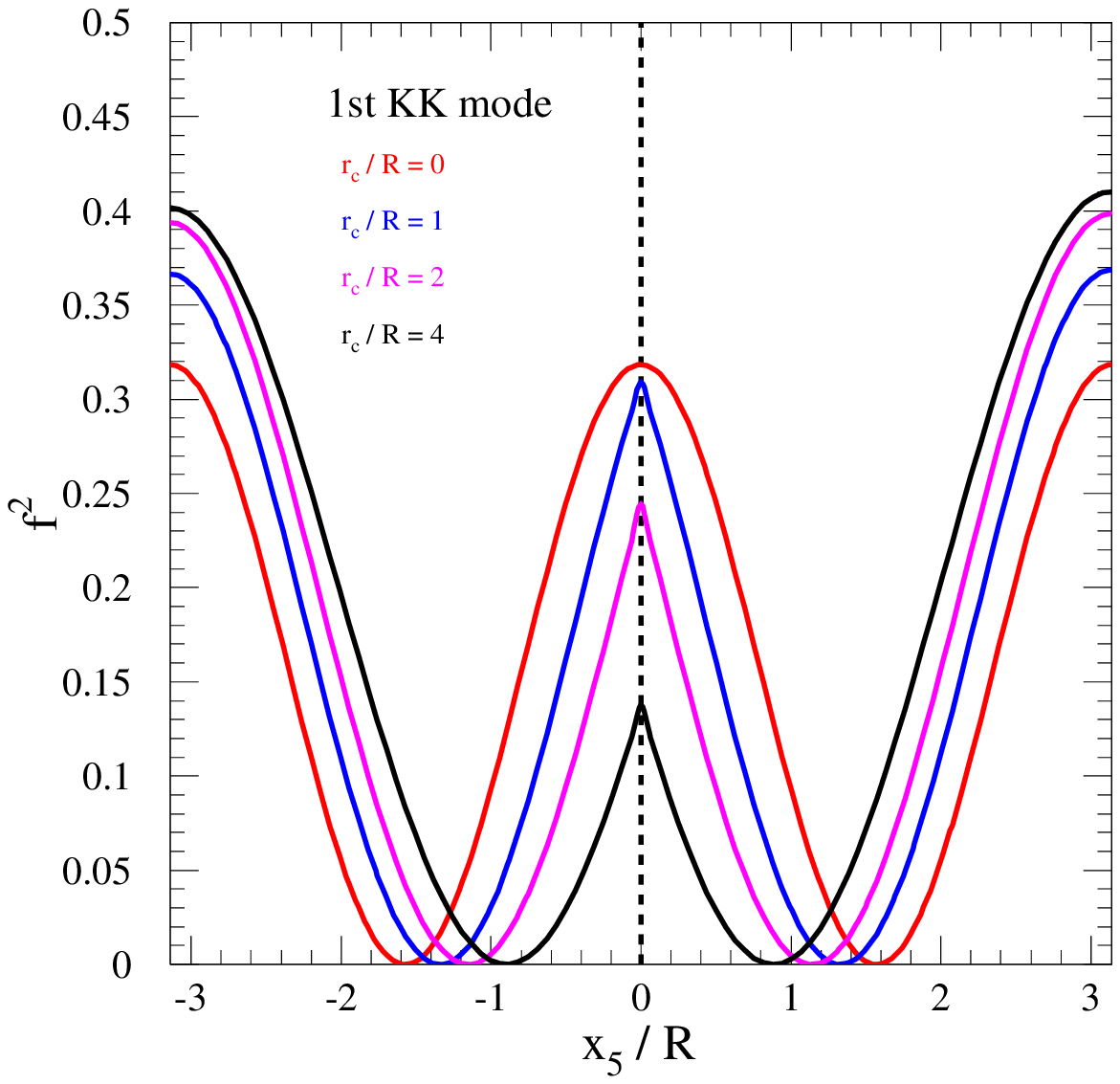} \hspace*{-1.5cm}
\epsfxsize=10.0cm\epsfysize=10.0cm
                     \epsfbox{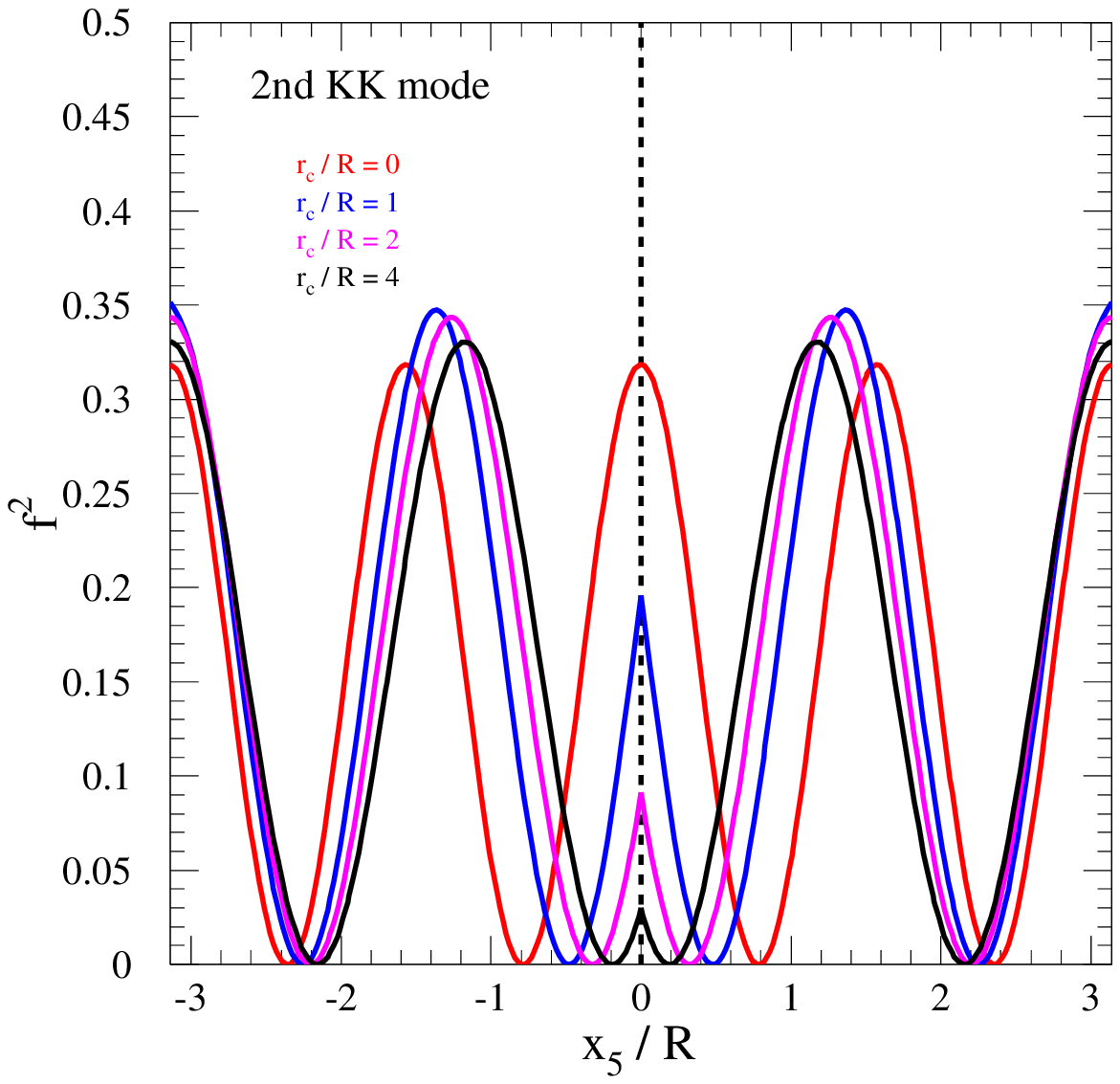}
}
\caption{The $n=1$ and $n=2$ KK mode wave functions 
for one brane with $r_c/R=0, 1, 2,$ and $4$.}
\label{fig:wf1}
\end{figure}

For the brane field at $x_\psi$ this results in coupling to the $n$th KK mode,
\bea
\label{eq:branecoupling}
\frac{ f_n( x_\psi ) }{\sqrt{Z_n}} .
\eea
The wave functions $f_n(x_5)$ for the first two modes are presented in
Figure~\ref{fig:wf1}.
Note that this implies that the zero mode gauge field, whose wave function
is constant, couples universally to all brane matter with
coupling,
\bea 
\frac{1}{g_0^2} &=& \frac{2 \pi R}{g_5^2} + \frac{1}{g_a^2}
\eea
irrespective of the location of the brane.  Of course, in principle a brane
containing charged matter located away from $x_5=0$ would also be opaque,
and should be included in our derivation of the $f_n(x_5)$.  We analyze this
case in the next sections.

For fields localized on the opaque brane itself, the relevant 
coupling to the higher modes may also be expressed,
\bea 
\frac{1}{g_n^2} &=& \frac{1}{f_n^2(0) g_5^2} + \frac{1}{g_a^2} .
\eea
In Figure~\ref{fig:masses} we show the ratio of $g_n^2$ to $g_0^2$
as a function of $r_c / R$.  Evident from the figure is the usual
$r_c / R = 0$ expectation that all modes couple equally strongly to the
brane fields, with coupling $g_n = \sqrt{2} g_0$.  However, once 
$m_n \sim 1 / r_c$ the brane no longer seems transparent, and the
KK modes have difficulty penetrating it,
decoupling from its fields.  This is evident from the wave functions 
(see Figure~\ref{fig:wf1})
themselves, which show significant distortion away from the brane once the
masses are greater than $1 / r_c$.

Couplings of the higher KK modes of bulk fields are model-dependent, 
being given by integrals of products of several of the wave functions.
For example, if there are bulk fermions, the coupling of the 
$n$-mode gauge field to two bulk fermion modes with wave functions 
$f^i_\psi (x_5)$ and $f^j_\psi (x_5)$ is,
\bea
\frac{1}{\sqrt{Z_n}} \int dx_5 \: f_n(x_5) \; {f^i_\psi (x_5)}^* \;
f^j_\psi (x_5)   .
\eea
Thanks to the orthonormality of the fermion KK 
decomposition,the zero mode gauge field couples only to two
fermions of the same mode number, with universal coupling $g_0$.
For the three- and four-point gauge field vertices, we have,
\bea
g_{nml} & = &
\frac{1}{\sqrt{Z_n Z_m Z_l}} \times \int dx_5 \: 
\left( \frac{1}{g_5^2} + \frac{\delta (x_5)}{g_a^2} \right) \:
\: f_n(x_5) \, f_m(x_5) \,  f_l(x_5) \\
g_{nmlk} & = &
\frac{1}{\sqrt{Z_n Z_m Z_l Z_k}} \times \int dx_5 \: 
\left( \frac{1}{g_5^2} +  \frac{\delta (x_5)}{g_a^2} \right) \:
f_n(x_5) \, f_m(x_5) \, f_l(x_5) \, f_k(x_5)
\eea
between the $A_n$-$A_m$-$A_l$ and $A_n$-$A_m$-$A_l$-$A_k$ modes,
respectively.  We have suppressed the vector and color indices, but these
are simply restored.  

The above results can be simplified for the vertices 
involving the zero mode, because its wave function is independent of 
$x_5$ and thus drops out of the integration.  In the three-point vertex,
we find that setting $l=0$ reduces the integration to the same one
which diagonalized the kinetic energy term; thus 
using Equation~(\ref{eq:diagonal}) the integration gives
$Z_n \delta_{m n}$ and the vertex factor is 
$f_0(0) \delta_{m n} / \sqrt{Z_0} = g_0 \delta_{m n}$ for all modes.
In the four-point vertex, setting $l=k=0$ results in the same integral,
$Z_n \delta_{m n}$, and the vertex factor is thus 
$f^2_0(0) \delta_{m n} / Z_0 = g^2_0 \delta_{m n}$.  Together, these
results demonstrate the fact that the zero mode gauge field's couplings
take a universal form as dictated by its unbroken gauge invariance, resulting
in the same coupling to both bulk and brane fields.

\subsection{Two Opaque Branes}

It is relatively simple to generalize our results to include two
branes at the orbifold fixed points, one at $x_5 = 0$ 
and one at $x_5 = \pi R$.  The action thus contains,
\bea
\label{eq:s5d2}
S &=& \int d^5 x \left\{
-\frac{1}{4 g_5^2} {\cal F}^{M N} {\cal F}_{M N} 
- \delta (x_5) \frac{1}{4 g_a^2} {\cal F}^{\mu \nu} {\cal F}_{\mu \nu}
- \delta (x_5- \pi R) \frac{1}{4 g_b^2} {\cal F}^{\mu \nu} {\cal F}_{\mu \nu} 
\right\} .
\eea
The wave functions satisfy the eigenvalue equation in terms
of $r_a \equiv g_5^2 / g_a^2$ and $r_b \equiv g_5^2 / g_b^2$,
\bea
\left[ \partial_5^2 + m_n^2 + r_a m_n^2 \delta (x_5) 
+ r_b m_n^2 \delta (x_5 - \pi R) \right] f_n & = & 0 .
\eea
Wave functions can be constructed by the same technique used in
the previous section to match solutions across the branes.  The
resulting solutions are,
\bea
\label{eq:eigenmodes2}
f_n (x_5) = {\cal N}_n \left\{ 
\begin{array}{lcr}
\cos [m_n x_5 ] + (m_n r_a / 2) \sin [m_n x_5] & & x_5 < 0 \\[0.2cm]
\cos [m_n x_5 ] - (m_n r_a / 2) \sin [m_n x_5] & & x_5 \geq 0
\end{array}
\right. ,
\eea
in which the dependence on $r_b$ is hidden inside the
quantized masses, which satisfy the relation,
\bea
\label{eq:eigenmass2}
0 & = & \left( r_a \, r_b \, m_n^2 - 4 \right) \tan [m_n \pi R] 
- 2 \, (r_a + r_b) \, m_n.
\eea
This equation may be reduced to the form $m_n = $ a function of
$\tan [m_n \pi R]$ using the quadratic equation, and solved
numerically as before.  This determines the mass spectrum
and couplings of the KK gauge bosons.  Continuing to normalize the
integral of $f^2_n$ to one, the $Z_n$ normalization factors are,
\bea
Z_n & = & \left( \frac{1}{g_5^2} + \frac{f_n^2(0)}{g_a^2} + 
\frac{f_n^2(\pi R)}{g_b^2} \right) .
\eea

The couplings of given modes or combinations of modes are derived
from these results.  For brane field couplings to a single KK mode, one
finds the same equation (\ref{eq:branecoupling})
as before in terms of the new wave functions.
The couplings among gauge modes are,
\bea
g_{nml} & = & \frac{1}{\sqrt{Z_n Z_m Z_l}} \times \int dx_5 \: 
\left( \frac{1}{g_5^2} + \frac{\delta (x_5)}{g_a^2} 
 + \frac{\delta (x_5-\pi R)}{g_b^2} \right) \:
\: f_n(x_5) \, f_m(x_5) \,  f_l(x_5) \\
g_{nmlk} & = &
\frac{1}{\sqrt{Z_n Z_m Z_l Z_k}} \times \int dx_5 \: 
\left( \frac{1}{g_5^2} +  \frac{\delta (x_5)}{g_a^2} 
 + \frac{\delta (x_5-\pi R)}{g_b^2} \right) \:
f_n(x_5) \, f_m(x_5) \, f_l(x_5) \, f_k(x_5)
\eea
In particular, the zero mode coupling is,
\bea
\frac{1}{g_0^2} & = & \frac{2 \pi R}{g_5^2} + \frac{1}{g_a^2} +
\frac{1}{g_b^2} .
\label{eq:twobraneg0}
\eea

Before considering specific two-brane configurations, we note that 
many of these formulae are easy to generalize.  In particular, the $Z_n$
generalize into an obvious sum of $1/g_5^2$ plus $f_n^2(x_i) / g_i^2$
for each opaque brane at $x_i$ with coefficient $1/g_i^2$.  The coupling to 
brane matter fields always takes the same form, and the bulk couplings 
generalize to an integral over the same product of the $f_n$ times 
$1/g_5^2$ plus $\delta(x_5-x_i)/g_i^2$.  What remains is to determine the 
eigenmass equation and associated wave functions for a given set of branes, 
a straight-forward (but in the case of many branes, tedious) exercise.

\subsubsection{Symmetric Branes}

\begin{figure}[t]
\vspace*{-1.cm}
\centerline{ \hspace*{1cm}
\epsfxsize=10.0cm\epsfysize=10.0cm
                     \epsfbox{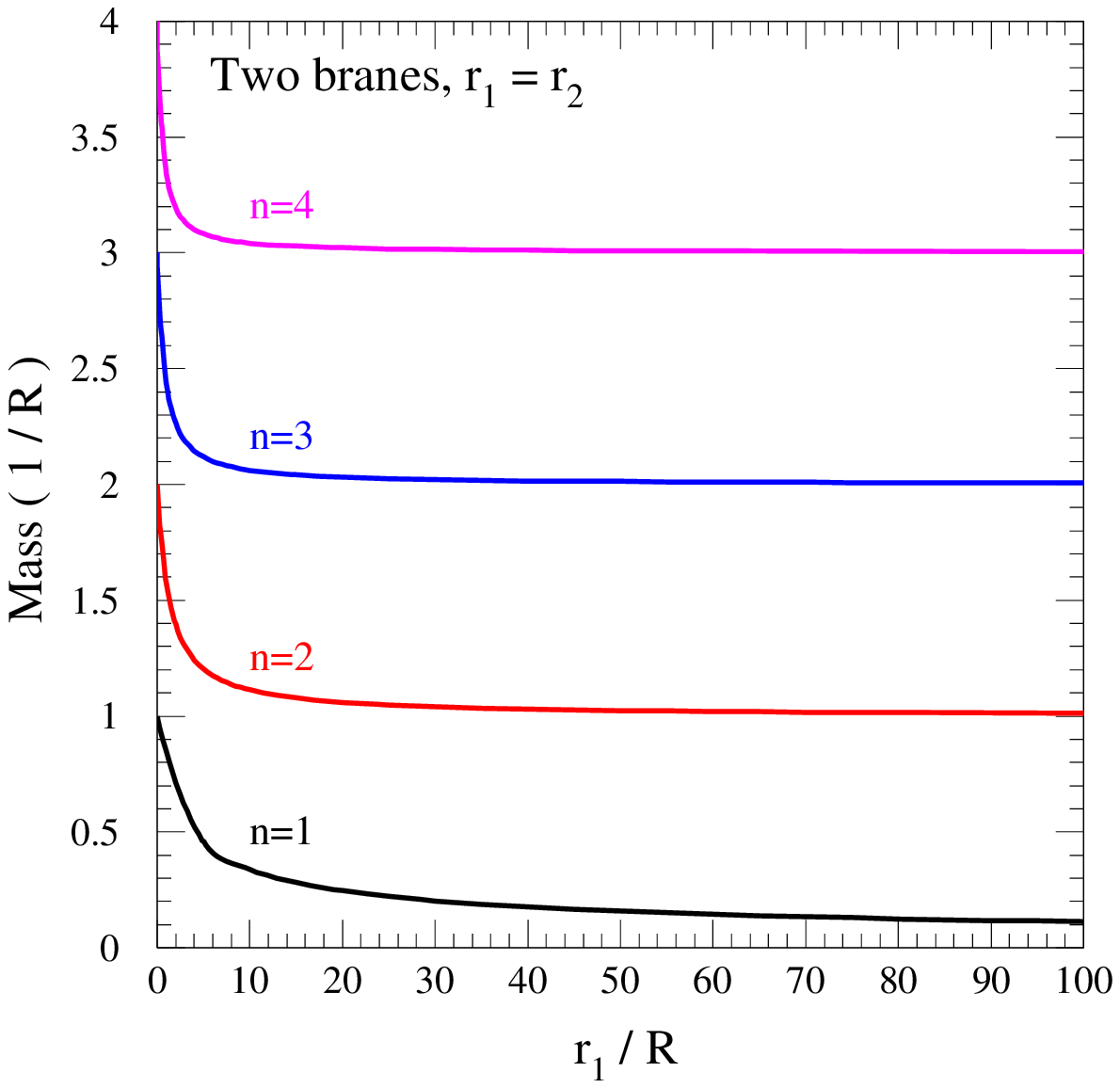} \hspace*{-1.5cm}
\epsfxsize=10.0cm\epsfysize=10.0cm
                     \epsfbox{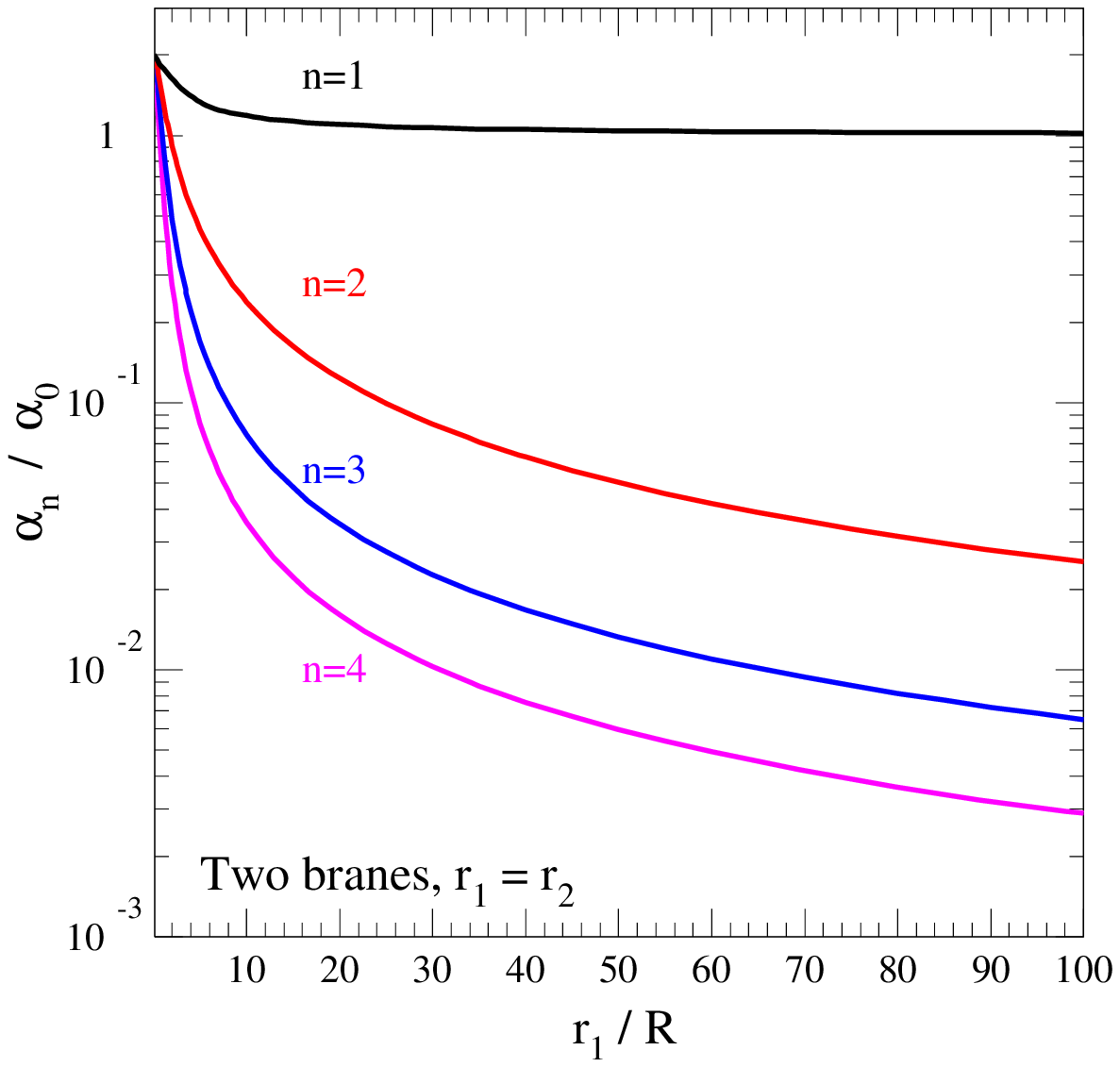}
}
\caption{The $n=1,2,3,4$ (bottom to top) KK mode masses in units of $1/R$
and KK mode couplings relative to the zero mode coupling
as a function of $r_c / R$ for two branes with equal terms.}
\label{fig:masses2}
\end{figure}

For our first example, consider equal brane kinetic terms,
$r_a = r_b \equiv r_c$.  This is the case induced by radiative corrections
to 5d theories with orbifold boundary conditions and no brane fields.  
Note that this set-up
preserves a $\Z_2$ symmetry under which even number KK modes are even and
odd number modes are odd.  This KK parity forces couplings involving
an odd number of odd-mode fields to vanish.
The resulting mass spectrum satisfies the equation,
\bea
\frac{r_c \, m_n}{2} & = & \frac{ \cos [m_n \pi R] \pm 1}
{\sin [m_n \pi R]}
\hspace*{2cm} \left( m_n \geq 0 \right) \: , 
\eea
and is shown for the first few KK modes in Figure~\ref{fig:masses2}.
The $+$ sign in the equation is realized for the first KK mode, and
higher modes are realized for alternating signs.  
An interesting feature of the two brane case is evident for 
$r_c / R \gg 1$, in which the mass of the
first KK mode approaches zero and the remaining modes approach their
canonical values of $(n-1) / R$.  This contrasts with the one brane case,
for which the first KK mode mass was always greater than $1 / 2 R$.
One can solve exactly
for a solution with $m_1 R \ll 1$ by expanding the $\sin$ and $\cos$
in the eigenmass equation.  The result is,
\bea
m^2_1 & = & \frac{4}{r_c^2} \left( 1 + \frac{r_c}{\pi R} \right) 
\approx \frac{4}{\pi \, r_c \, R} .
\eea
This lightest KK mode can
be understood to be a sort of ``collective mode'', composed
of a tiny amount of every wave function in the otherwise unperturbed tower.  
The coupling of the $n$-mode gauge fields to
fields confined on either opaque brane may be expressed as,
\bea
\frac{1}{g_n^2} & = & \frac{1}{g_5^2 f_n^2(0)} + \frac{2}{g_a^2}
\eea
and is plotted for the first few modes in Figure~\ref{fig:masses2}.
Unlike the rest of
the tower, which exhibits similar decoupling from the branes seen in the
one brane case, the collective mode's coupling approaches the zero mode
coupling in the limit of large $r_c/R$.
We have chosen to extend Figure~4 up to $r_c / R= 100$ in order to display the 
asymptotic behavior as a function of $r_c/R$.  

\begin{figure}[t]
\vspace*{-1.cm}
\centerline{ \hspace*{1cm}
\epsfxsize=10.0cm\epsfysize=10.0cm
                     \epsfbox{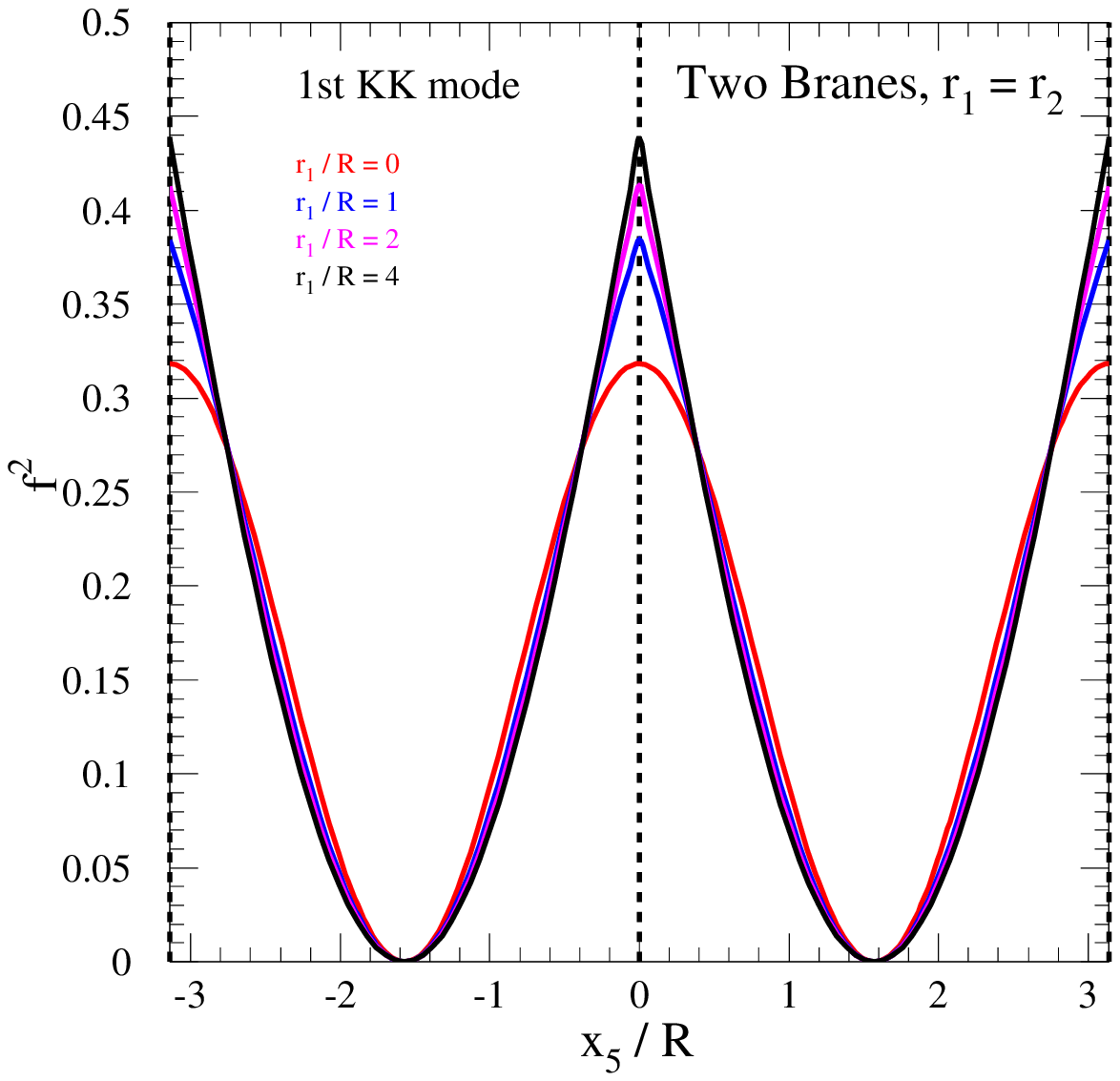} \hspace*{-1.5cm}
\epsfxsize=10.0cm\epsfysize=10.0cm
                     \epsfbox{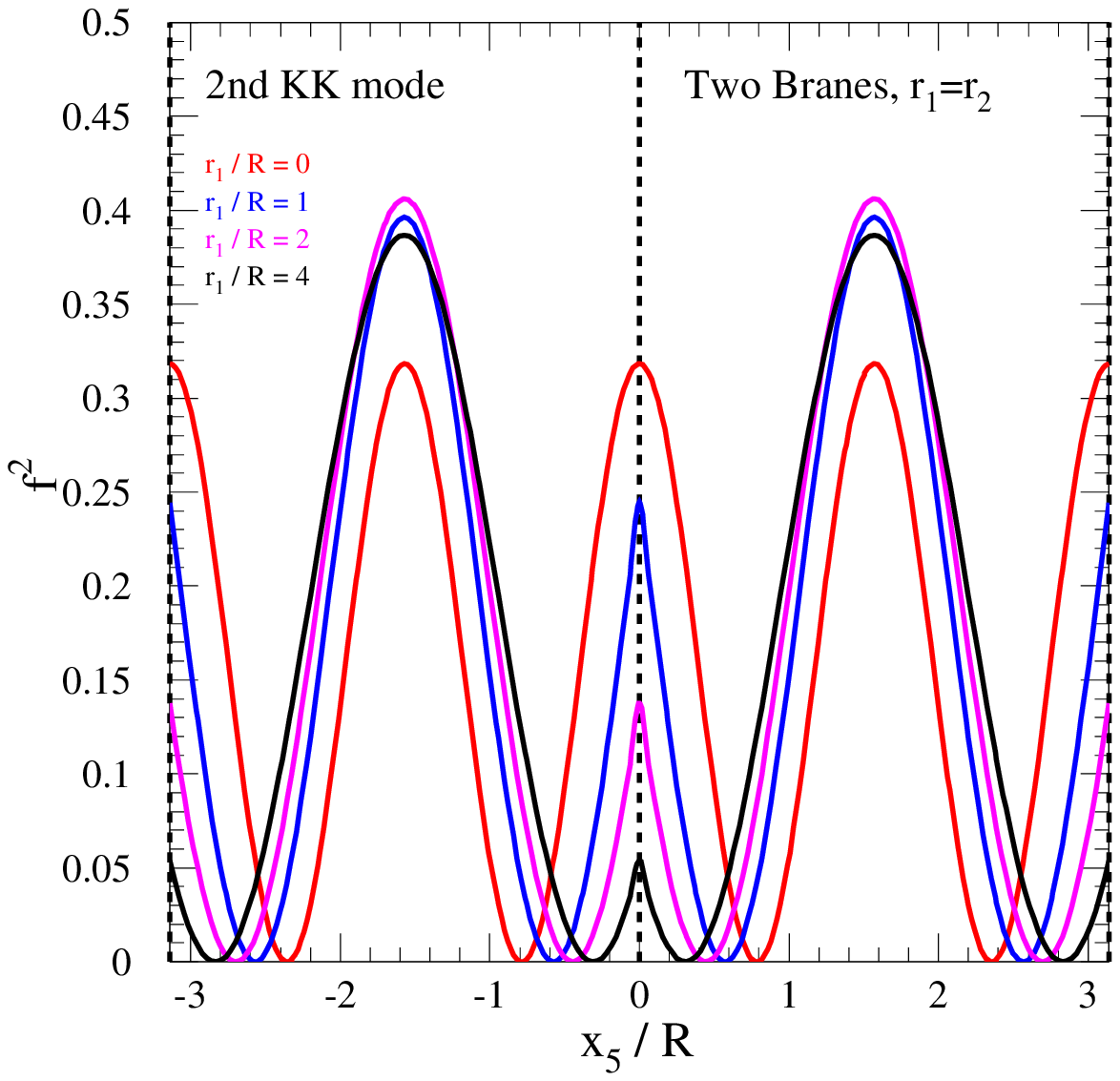}
}
\caption{The $n=1$ and $n=2$ KK mode wave functions 
for two branes with equal $r_1/R=0, 1, 2,$ and $4$.}
\label{fig:wf2}
\end{figure}

One can gain intuition into the reason why this feature appears in the
case with two opaque branes by considering an observer living on the brane
at $x_5=0$ and measuring the gauge coupling at that brane by scattering
various types of charged matter at various energies.
At distances somewhat shorter than the size of the extra dimension, the 
observer fails to realize that the other brane is present, and scattering
between matter localized on different branes should cease.  Furthermore,
at these energies the distances probed are too short to realize that
there is a second opaque brane at all, and the observed coupling
should not depend on $g_b$.  At very high energies, the distances probed
are short enough that the fact that there is an extra dimension becomes
irrelevant and the coupling should be dominated by the coupling present on the 
brane, and thus must approach $g_a$.  However, in contrast to the
one brane case, the zero mode coupling does not approach $g_a$, but to
a combination of $g_a$ and $g_b$.  Thus, something is needed to restore the
correct behavior, and the higher KK modes will not serve because they
decouple from the brane.  Thus, the collective mode's couplings must
approach the zero mode coupling (for $g_a = g_b$) with an appropriate
relative sign in order for the net force to be described by $g_a$ alone.

\begin{figure}[t]
\vspace*{-1.cm}
\centerline{
\epsfxsize=12.0cm\epsfysize=12.0cm
                     \epsfbox{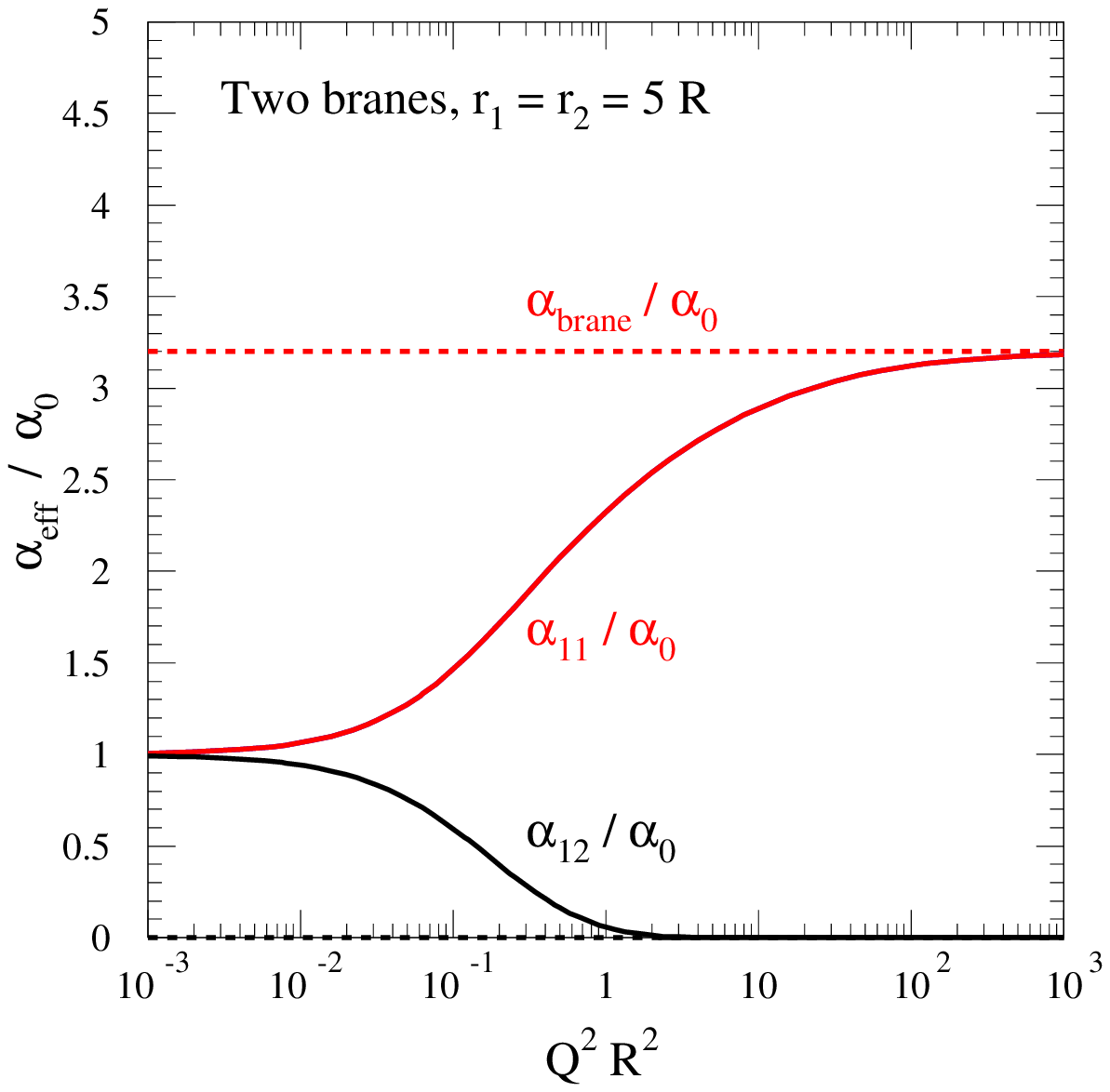}
}
\caption{The effective coupling between two fields on the same brane
(upper line) and two fields on different branes (lower line) as
a function of momentum transfer, $Q^2$.}
\label{fig:geff1}
\end{figure}

One can further explore this intuitive picture by examining the effective
coupling between fields either on the same brane or on different branes.
In the KK description of the theory, the net force between them is a
sum over all of the KK modes,
\bea
\frac{1}{4 \pi} \sum_{n \geq 0} \frac{g^i_n g^j_n}{Q^2 + m_n^2} 
= \frac{\alpha_0}{Q^2} 
\left( 1 + \sum_{n \geq 1} \frac{g^i_n g^j_n / \alpha_0}
{1 + m_n^2 / Q^2}  \right) \ra \frac{\alpha_{ij} (Q^2) }{Q^2}
\eea
where $Q^2$ is the momentum transfer, $g^i_n$ is the coupling of the
$n$th KK mode to the $i$-th brane, and $\alpha_{ij} (Q^2)$ is an effective
coupling which includes the exchange of all KK modes in the interaction
of brane field $i$ with brane field $j$.  Using our numerical solution
for the symmetric two brane case, we can explicitly compute the effective
intra- and inter-brane couplings.  The result is shown in 
Figure~\ref{fig:geff1}, and illustrates the physics described above.
At low $Q^2$ the exchange is dominated by the zero mode, and the two
effective couplings are equal to the zero mode coupling.  The effect of
the collective mode appears rather early, thanks to its small mass, and
the couplings begin to differ.  For $Q^2 \gsim 1/ \pi^2 R^2$, interactions
occur at distance scales smaller than the separation between the two branes,
and the intra-brane coupling vanishes as each brane fails to realize that
the other is there.  Finally, at very large momentum transfer the brane
field fails to realize that there is an extra dimension, and the physics
is described entirely by its own (four dimensional) gauge term
with coupling $g_a$.

\subsubsection{Asymmetric Branes}
\label{sec:asym}

\begin{figure}[t]
\vspace*{-1.cm}
\centerline{ \hspace*{1cm}
\epsfxsize=10.0cm\epsfysize=10.0cm
                     \epsfbox{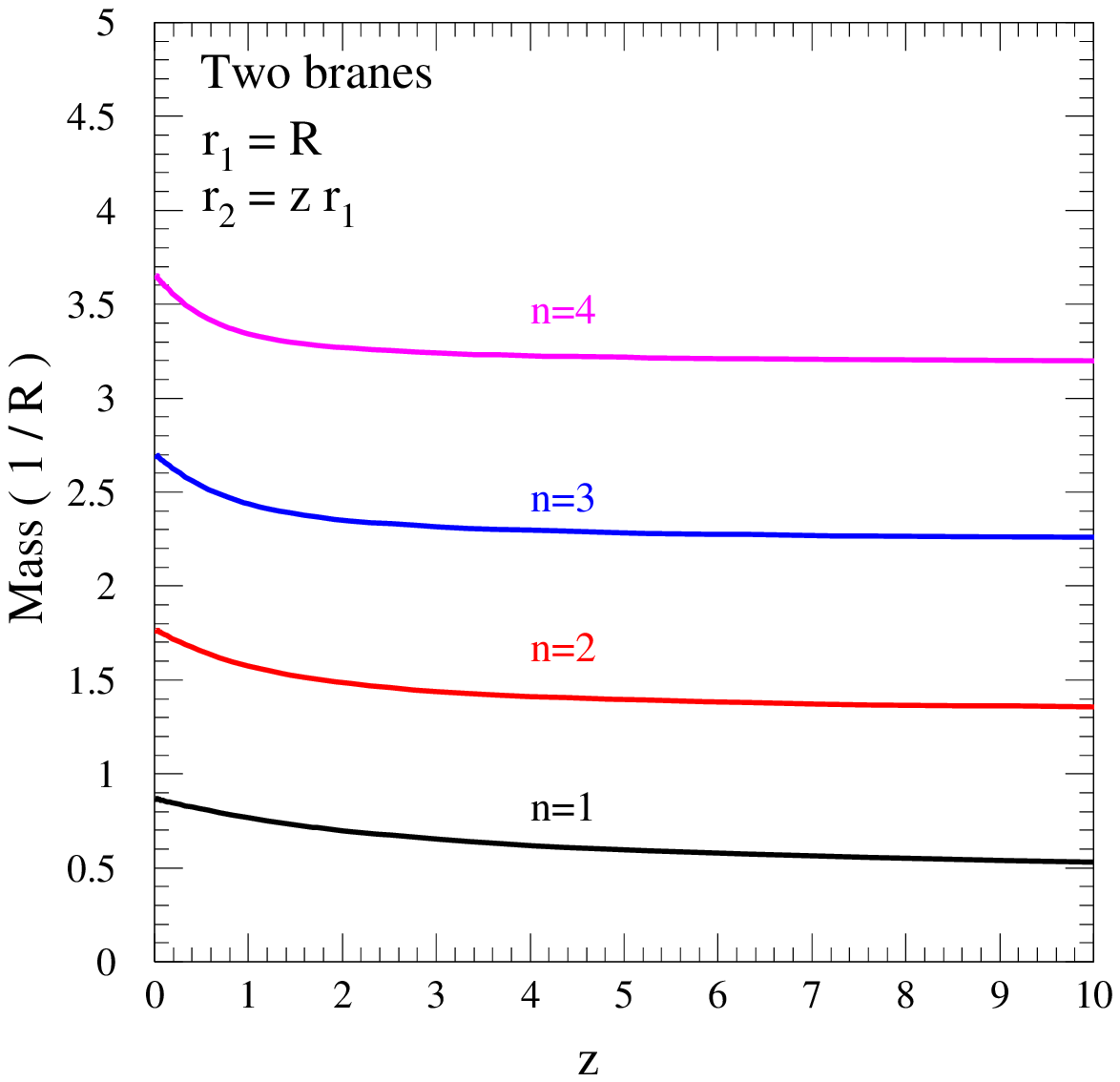} \hspace*{-1.5cm}
\epsfxsize=10.0cm\epsfysize=10.0cm
                     \epsfbox{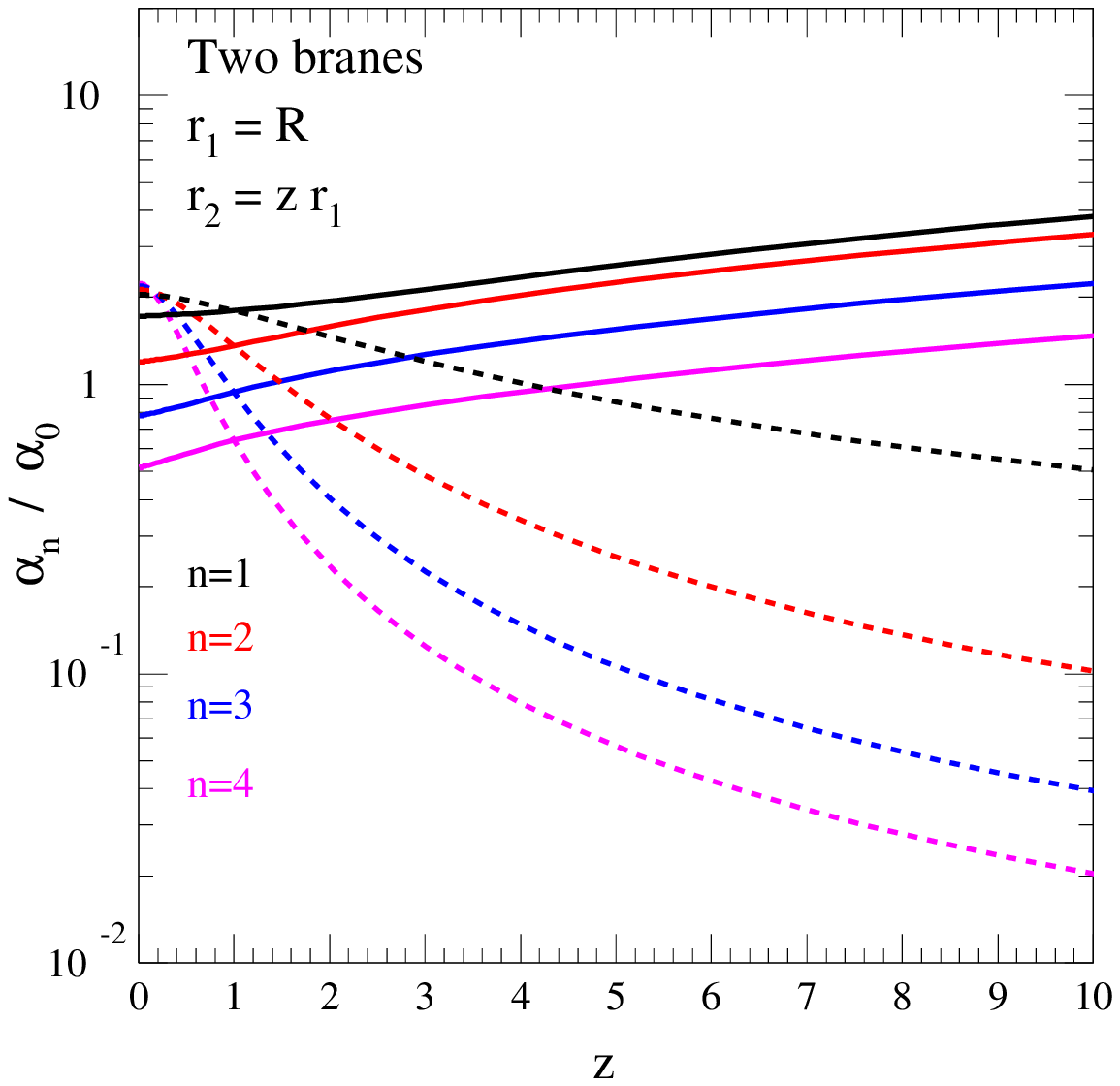}
}
\caption{The $n=1,2,3,4$ (bottom to top) KK mode masses in units of $1/R$
and KK mode couplings relative to the zero mode coupling
(solid lines are the couplings to brane fields
at $x_5=0$ and dashed lines are couplings to brane fields at $x_5=\pi R$)
as a function of $z$ for two branes with kinetic terms
$r_1=R$ and $r_2 = z r_1$.}
\label{fig:masses3}
\end{figure}

Another interesting case has two branes with different terms.
For simplicity, we fix the term on the first brane (at $x_5=0$) 
to $r_a = r_c$ and allow the term on the second brane (at $x_5=\pi R$)
to vary as $r_b = z r_a$.  For $z \not = 1$, this configuration
explicitly violates KK parity.  As motivation, one might imagine a 
construction in which some fermions (for example, the leptons) are 
confined to the brane at $x_5=0$ while some others (for instance, 
the quarks) are confined to the other brane at $x_5 = \pi R$.  This
configuration can suppress local operators leading to unacceptably
fast proton decay because of a low Planck scale \cite{Arkani-Hamed:1999dc}.
Given the asymmetry between the two branes, it would be somewhat contrived
if the gauge kinetic terms living on them were the same.  In addition,
the much larger number of quark degrees of freedom will imply very
different quantum corrections to both terms, so the choice of equal
couplings at the two branes is only justified in some particular cases.

\begin{figure}[t]
\vspace*{-1.cm}
\centerline{ \hspace*{1cm}
\epsfxsize=10.0cm\epsfysize=10.0cm
                     \epsfbox{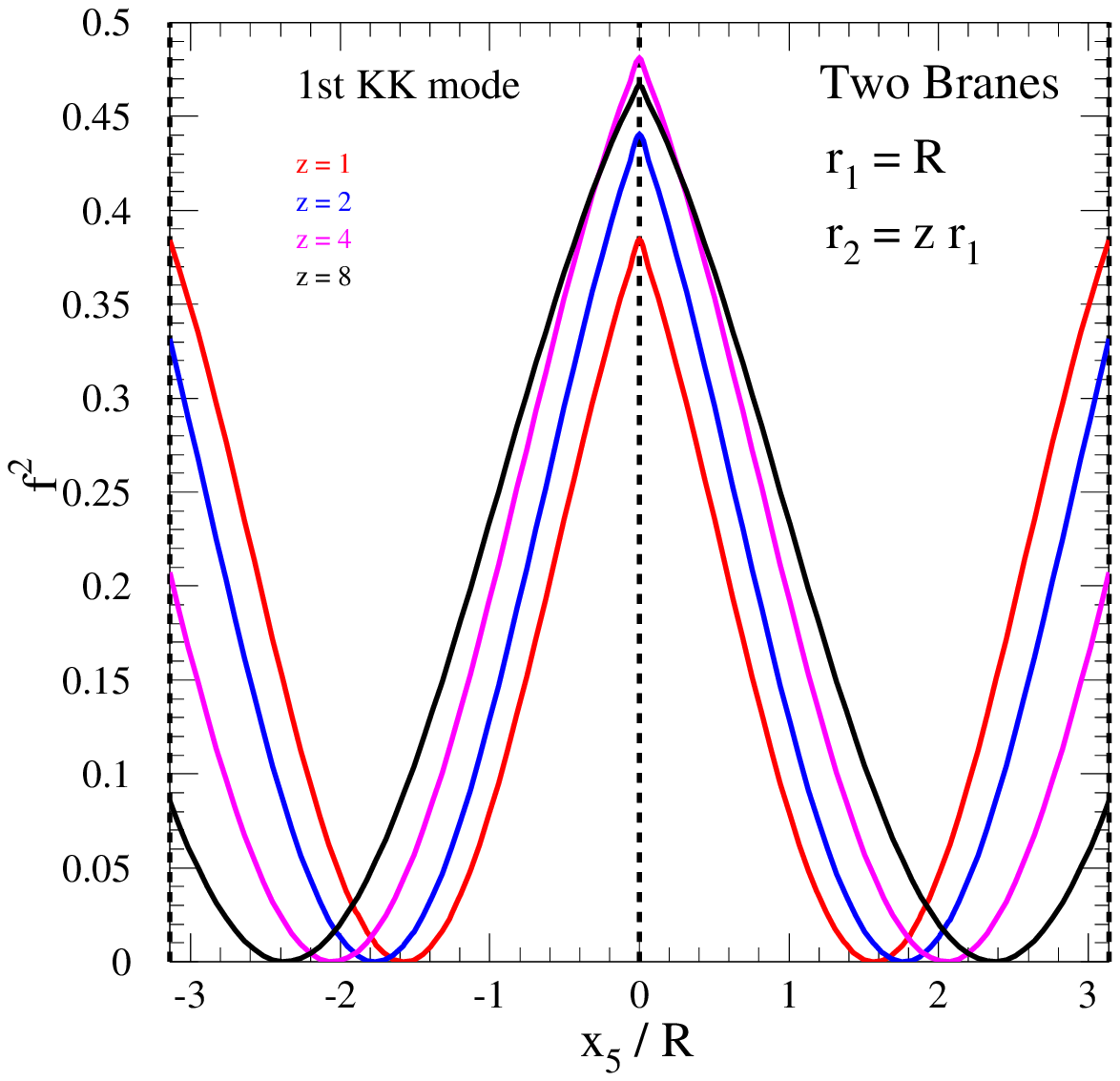} \hspace*{-1.5cm}
\epsfxsize=10.0cm\epsfysize=10.0cm
                     \epsfbox{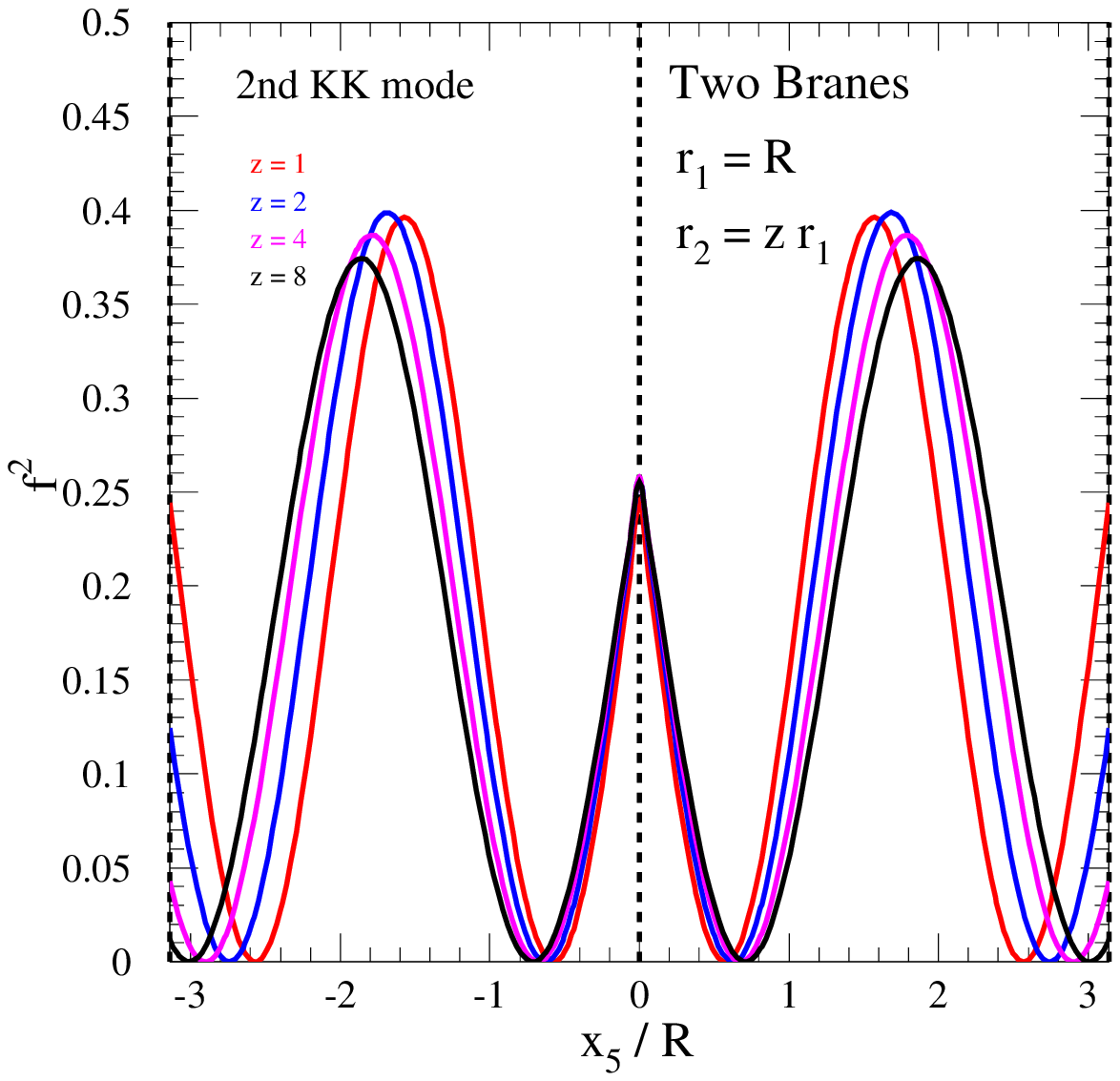}
}
\caption{The $n=1$ and $n=2$ KK mode wave functions 
for two branes with $r_1/R=1$ and $r_2 = z r_1$ for $z=1,2,4,8$.}
\label{fig:wf3}
\end{figure}

The KK masses are now solutions of the general two brane eigenmass
equation~(\ref{eq:eigenmass2}).  In terms of $z$ and $r_1$ its 
solutions can be expressed,
\bea
m_n r_1 & = & 
\frac{ (1+z) \pm \sqrt{(1+z)^2 + 4 \tan^2 \left[ m_n \pi R \right]}}
{z \tan \left[ m_n \pi R \right]} ,
\eea
where again the $+$ sign is realized for the first KK mode solution,
and successive modes are realized for alternating signs.
Results are plotted for $r_1 = R$
and $r_2 = z R$ in Figure~\ref{fig:masses3}.  The features are roughly
similar to those evident in the symmetric two brane case, including
the existence of a collective mode in the limit of $r_1, r_2 \gg R$ 
with very small mass,
\bea
m^2_1 & = & \frac{2}{z r_1^2} \left( 2 + \frac{r_1 (1+z)}{\pi R} \right) 
\approx \frac{2(1+z)}{z \pi \, r_1 \, R} .
\eea

Having found the masses, the next step is to examine the wave functions.
We expect that for asymmetric branes, the larger brane term will dominate,
pushing the wave functions further away from that brane.  This implies that
the higher KK modes couple more weakly to brane fields on one of the opaque
branes than to those fields on the other.  This is evident in the wave
functions, plotted for the first two modes in Figure~\ref{fig:wf3}.
For $z=1$ we see, apart from an over-all sign, 
equal coupling at both branes, whereas for $z \gg 1$
the wave functions become very small on the brane at $x_5 = \pi R$.
This has the implication that the KK modes couple much more strongly
to fields located on the less opaque brane than to fields localized on
the more opaque brane.  
We plot these couplings in Figure~\ref{fig:masses3},
and find that the effect is quite striking for $z \gg 1$.
One can understand this feature in terms of the
effective couplings on each brane.  $z \gg 1$ implies that $g_b^2$ is
very small, and thus dominates the zero mode coupling.  This in turn
implies that, since the effective coupling in the ultra-violet should
converge to the local brane terms, KK modes should rapidly decouple from
the second brane, while they must couple relatively strongly to the first
one in order to make up the difference between the zero mode coupling
and the local term on that brane.

%%%%%%%%%%%%%%%%%%%%%%%%%%%%%%%%%%%%%%%%%%%%%%%%%%%%%%%%%%%%%%%
\section{Implications for Phenomenology}
\label{sec:pheno}
%%%%%%%%%%%%%%%%%%%%%%%%%%%%%%%%%%%%%%%%%%%%%%%%%%%%%%%%%%%%%%%

Our results can have profound implications for the phenomenology of
models in which gauge fields propagate in the bulk.  The standard picture
for this situation is that high energy colliders can identify an extra 
dimension by discovering the tower of KK modes with masses $n/R$ and
couplings $\sqrt{2}$ times greater than the zero mode coupling to 
brane-localized matter \cite{Rizzo:1999br,Lykken:1999ms}.  In the limit of
a very small gauge-kinetic term on the brane these results will approximately
hold for a large number of KK modes, and this phenomenological picture will
remain valid.  However, the results above indicate that these theories
have, in addition to the parameter $R$ which controls the size of the extra
dimension and thus the masses of the KK modes, at least 
one other, ``hidden'' 
parameter, $r_c$, which will distort the KK spectrum and modify the
couplings to brane fields.  Since it is effectively a tree level coupling,
it is somewhat arbitrary to set it equal to zero.

\subsection{One Brane Case}

As a simple example of what the local brane couplings may do to limits
from colliders, let us consider
the effect of virtual KK photon and $Z$ exchange on the process
$e^+ e^- \ra f \overline{f}$.  
This was discussed for the transparent brane case in
Ref.~\cite{Rizzo:1999br}.
At energies far below the mass of the
first KK mode, the effect of the virtual KK photon exchange can be included 
model-independently as a four-fermion operator,
\bea
{\cal O}_\gamma & = & -e^2 Q_e Q_f \frac{V}{m_W^2} 
\left[ \overline{e} \gamma^\mu e \right]
\left[ \overline{f} \gamma_\mu f \right] ,
\eea
where the dependence on $R$ is hidden inside the coefficient $V$,
\bea
V & = & m_W^2 \: \sum_{n \geq 1} \, \frac{\alpha_n / \alpha_0}{M_n^2} ,
\label{eq:Vnew}
\eea
where $M_n$ is the mass of the $n$th KK gauge boson and $\alpha_n$ the
product of its couplings to the $e$ and $f$ fields.

In the transparent brane case,
\bea
V & = & m_W^2 R^2 \: \sum_{n \geq 1} \; \frac{2}{n^2}
\: = \: 2 m_W^2 R^2 \zeta (2)
\hspace*{4cm} (r_c = 0),
\label{eq:Vold}
\eea
with the factor of $2$ a result of the KK mode couplings being
$\sqrt{2}$ times the zero mode couplings, and the sum over $n$
includes all KK states, whose masses are $n/R$.
In the five dimensional case we consider the sum is convergent to
$\zeta(2)$ as indicated (though the result changes only by about $5\%$
if truncated at $n=10$),
but in higher dimensions it would have to be cut-off in some fashion,
introducing dependence on the UV completion.
Similar operators can be written for KK modes of the $W^\pm$ and $Z$
(and in the case of ${\cal O}_Z$ will generally interfere with 
${\cal O}_\gamma$ in physical processes).  In the $r_c \ra 0$
limit, the quantity $V$ will be the same for all
of the bulk gauge fields.  Thus, including all relevant operators,
Ref.~\cite{Rizzo:1999br} deduces that with $200~{\rm pb}^{-1}$ of 
$\sqrt{s}=195$ GeV LEP data the reach extends up to
$V \lsim 4.5 \times 10^{-3}$ which corresponds to 
$R^{-1} \lsim 2.2$ TeV.  At the NLC with $\sqrt{s}=500$ GeV and collecting
$500~{\rm fb}^{-1}$ of data the bound becomes $V \lsim 1.2 \times 10^{-4}$
or $R^{-1} \lsim 13$ TeV.

The situation with $r_c \sim R$ can be quite different.  For example,
we consider the one-brane case with both $e$ and
$f$ fields living on the opaque brane.  To compute $V$ one must return to the
definition, Eq.~(\ref{eq:Vnew}) where $\alpha_n$ and $M_n$ are now 
complicated functions of $R$ and $r_c$,
as shown in Figure~\ref{fig:masses}.  The leading term (from the first
KK mode) may be somewhat enhanced by the mass of that mode being lighter
than $1/R$, but is also somewhat decreased by the suppressed coupling to
brane fields.  The suppression dominates the enhancement.
The higher KK number states are still approximately
equally spaced in mass, but their couplings to brane fields 
become highly suppressed, and the sum in Equation~\ref{eq:Vnew} converges
much more quickly than the $r_c =0$ expression in Equation~\ref{eq:Vold}.
Using the new expression for $V$, we translate the LEP and NLC bounds
on $V$ (which are independent of the new physics) into the plane of
$1/R$ and $r_c/R$.  For $r_c / R \gsim 1$ the limits on $1/R$ can
be substantially modified; for the LEP (projected NLC) limits
derived above, we have, for $r_c / R \sim 1$, $R^{-1} \gsim 2$ TeV
(12 TeV) and for $r_c / R \sim 10$, $R^{-1} \gsim 1.3$ TeV (8.1 TeV).

Another interesting possibility is the fact that each gauge group in the bulk
may have a separate $r_c$ on the brane.  This would allow mass splittings
between the KK modes much larger than one would normally consider from
radiative corrections.  This allows each gauge boson
to have its own $V$, which would not be expected from the simple 
extra-dimensional picture with transparent branes.  It further has the effect 
that the KK modes for the neutral weak boson sector can have a different 
Weinberg angle than the one observed for the zero modes, and thus may be
poorly approximated as KK modes of the ordinary photon and $Z$, 
and better represented as different mixtures of heavy copies of 
the SU(2) and U(1) neutral bosons.  A similar effect can also occur
when some of the weak gauge groups and/or Higgs fields are confined to
a brane \cite{Muck:2001yv}.

\subsection{Two Branes and Split Fermions}

When there are two or more branes, the lightest KK mode is generally
a collective mode which does not decouple from the brane, and whose mass is 
not characterized by $R^{-1}$, but instead by $1/ \sqrt{ R r_c }$.  
This leads to the interesting possibility in which a first KK mode can be 
discovered, but higher modes (whose masses are characterized by $R^{-1}$ 
and whose couplings to brane fields are suppressed) remain out of reach.
This is very different from the transparent brane case in which one
expects evenly spaced KK modes, so that the second mode has mass twice as 
large as the first mode and the same coupling strength.
The discovery potential for the collective mode will be similar to
existing searches for standard $W^\prime$ and $Z^\prime$ bosons, 
colorons \cite{Pillai:1996st}, and so forth.

In models with low-scale quantum gravity, there is motivation to 
consider the possibility that quarks and leptons live on separate 
branes\footnote{Separating fermions in the extra dimension may introduce
local anomalies which may be canceled by a Chern-Simons term 
\cite{Arkani-Hamed:2001is}.  Such a term has no affect at the perturbative 
level.}, 
in order to prevent dangerous operators which would mediate proton decay
in conflict with existing bounds \cite{Arkani-Hamed:1999dc}.  The
simplest implementation of such a picture has two branes, one containing
the quarks (and possibly the gluons) and the other containing the leptons.
One is thus forced to consider the weak gauge bosons propagating in the
bulk, and loops of the brane fermions should induce kinetic terms for
the weak gauge fields localized on each brane.  
Given the obvious asymmetry in the underlying dynamics which localized the
quarks on one brane but the leptons on the other, it seems natural that
one brane ({\em i.e.} the quark brane) could have a 
larger kinetic term than the other one ({\em i.e.} the lepton brane), 
and the results of 
Section~\ref{sec:asym} could be relevant to the phenomenology of the
KK modes.  This leads to two interesting variations on the usual phenomenology
of bulk gauge fields. The first is that, owing to the larger repulsion from
the quark brane than from the lepton brane, the KK modes may couple more 
weakly to quarks than to leptons.  This would alter the expected
relative production cross sections at, say high energy $e^+ e^-$ colliders and
hadron colliders, and would further affect the branching ratios into
a given species of fermion.  Furthermore, at large momentum transfers,
the two branes lose sight of each other, and at very high energies,
quarks and leptons {\em miss} each other because of their separation
in the extra dimension \cite{Arkani-Hamed:1999za}.  This is evident in
our two-brane results for the effective coupling between fields located
one on each of the two branes (see Figure~\ref{fig:geff1})
which approaches zero at high $Q^2$.
In contrast, quark-quark and lepton-lepton interactions remain
appreciable even at large $Q^2$.

Models which separate not only quarks from leptons, but also left-
and right-handed quarks and leptons from each other, may naturally
explain the observed hierarchy of fermion masses
\cite{Arkani-Hamed:1999dc,Dvali:2000ha,Kaplan:2001ga} by generating
small Yukawa couplings for the zero modes as the tiny overlap in
fermion wave functions.  Each localized fermion demands
a renormalization of the gauge field whose shape is related
to the profile of the fermion KK modes, and a full theory of flavor
could have as many separate contributions as there are fermions.
The resulting picture is therefore rather complicated and model-dependent,
and is beyond the scope of this work, but we can divine some general 
features from the simple cases we have studied.

First, we would see the high energy suppression of cross sections outlined
above for split quarks and leptons for any two different fermions, 
including same flavor fermions with different helicities!  Since the induced
localized gauge kinetic terms are sensitive to the shape of the fermion
zero mode wave functions, production cross sections and decay
branching ratios could be flavor-dependent in a complicated way.  The
properties of the KK gauge bosons thus provide one with a powerful
test of extra-dimensional flavor dynamics, exploring the cartography of
the extra dimension \cite{Rizzo:2001cy}.

Second, models with split fermions have strong flavor-changing neutral current
(FCNC) constraints from Kaluza-Klein modes of gauge fields
\cite{Kaplan:2001ga,Carone:1999nz,Delgado:1999sv}, because while the
gauge fields couple flavor-diagonally, the KK modes couple 
flavor-dependently, inducing FCNC's after the CKM rotation from the
gauge to mass basis is performed.  The limits derived in this way
on $R^{-1}$ from the Kaon system are quite strong, of order
$R^{-1} \gsim 100-1000$ TeV.  However, these limits could be relaxed 
quite substantially if appreciable $r_c$ terms are included.  Such 
terms will force the KK modes of the gauge fields to try to avoid the 
places where fermions are localized, and would limit the strength 
of the FCNC's.

%%%%%%%%%%%%%%%%%%%%%%%%%%%%%%%%%%%%%%%%%%%%%%%%%%%%%%%%%%%%%%%
\section{Conclusions}
\label{sec:conclusion}
%%%%%%%%%%%%%%%%%%%%%%%%%%%%%%%%%%%%%%%%%%%%%%%%%%%%%%%%%%%%%%%

Theories with extra dimensions offer both unique solutions to the puzzles
of particle physics as well as unique theoretical challenges.  To date, all
known descriptions must be regarded as effective theories, and without
a deeper model to describe the underlying physics responsible for the
compactification of dimensions, generation of branes and boundary
conditions, and confinement of fields
to brane world-volumes, the best one can do is to write down effective
descriptions which are self-consistent.  The theoretical motivations are
many, and the resulting phenomenology intriguing.  

We have explored a simple consequence of any theory with gauge fields in
the bulk of the extra dimensions, and charged matter either in the bulk
subject to orbifold boundary conditions, or confined to a brane.  Radiative
corrections to these theories mandate that such branes or boundaries are
not transparent to the gauge fields - instead they are opaque.  While the
opacity of the brane, parameterized by the size of a kinetic term for the
gauge field living on the brane or boundary, is not calculable in terms of 
other parameters in the theory without introducing assumptions about 
the nature of the UV completion, this does not justify ignoring it.  
Such terms may very well be large, and comprise an important part of 
relating a theory with extra dimensions to the real world.  The effect on 
phenomenology can be sizable, and the result qualitatively different from 
the situation in which they are neglected.  For example, charged fields 
confined to an opaque brane will decouple from the high Kaluza Klein modes 
of the gauge field, contrasting with the standard picture under with all 
KK modes couple equally to brane fields.  
This decoupling of the higher KK modes is somewhat similar to the effect of
dimensional deconstruction \cite{Arkani-Hamed:2001ca,Hill:2000mu}, 
which replaces an extra dimension with a chain of 4d gauge theories linked by 
scalar fields.  In the deconstructed case, the 
analogues of the KK modes of bulk fields
naturally distort and terminate at some high energy scale.  In the case of the
brane kinetic term, the KK spectrum remains infinite, but nevertheless
the coupling to brane fields becomes arbitrarily small for arbitrarily
heavy modes.

In addition, the existence of local gauge kinetic terms implies there may be 
collective KK modes whose masses are not related to the size of the 
extra dimension, but instead to the size of the brane kinetic term.
These collective modes typically do not decouple the way the higher
KK modes do, and thus have unique phenomenology compared with the
typical expectations of extra dimensional theories.  Finally,
theories which have different types of matter living at different locations
in an extra dimension can show the interesting behavior that at very
high energies interactions between different particles are suppressed.
At very high energies, the particles {\em miss} each other because of
their extra-dimensional separation.

In this article we have considered the gauge couplings at tree level.
For a complete analysis it would be necessary to evaluate
the renormalization group evolution of the couplings.  We reserve this
for future work, but comment on the salient features here.
For the zero mode gauge boson, the coupling will evolve approximately as 
expected in a five dimensional theory with transparent branes
\cite{Dienes:1998vh} because the zero mode couples universally to all fields.  
However, for non-Abelian gauge theories, there will be alterations 
to the evolution because of the shift in the positions of the thresholds 
of the higher KK modes.  However, the mass spacing between modes
of $1/R$ remains approximately valid.
For the higher KK mode fields, the evolution will be modified by the 
presence of the brane term.  For example, higher modes will barely feel 
any contribution to their running from fields on the opaque brane, and the 
contributions from other KK mode gauge bosons will be modified by the
distortion of the wave functions.

Our framework has been five dimensional theories with gauge fields living
in all five dimensions.  We have chosen this framework because it is
simple and predictive, but there are many alternatives to explore.  Our
results are representative for {\em any} bulk field, and suggest that
a complete, self-consistent effective theory including compact dimensions
has a few more parameters than one might naively guess.  The appearance of
divergences, which must be renormalized, implies that it is more generic
to treat these effects as tree-level couplings, in contrast to the naive
expectation that they arise as loop effects.  It would also be interesting
to explore larger numbers of compact dimensions, to see the results
in theories with six or more dimensions.  In addition, 
it would be exciting
to see if our results could be exploited in model-building, allowing
new extra dimensional theories to better explain the puzzles of the
Standard Model.

~\\
{\Large \bf Acknowledgements}\\
~\\
The authors are grateful for conversations with 
I.~Antoniadis, N.~Arkani-Hamed, C.~Csaki, E.~Dudas, G.~Gabadadze,
D.E.~Kaplan, M.~Peskin, A.~Pilaftsis, and S.~Pokorski.
Work at ANL is supported in part by the US DOE, Div.\ of HEP,
Contract W-31-109-ENG-38.

\end{document}